\documentclass[12pt, draftclsnofoot, onecolumn]{IEEEtran}
\usepackage[T1]{fontenc}
\usepackage{psfrag,amsmath,amssymb,color,cite,graphicx}
\usepackage{mathtools}
\allowdisplaybreaks
\usepackage{stfloats}
\usepackage[caption=false,font=footnotesize]{subfig}
\usepackage{algorithm}
\usepackage{mathrsfs}
\usepackage[algo2e,ruled,vlined]{algorithm2e} 
\usepackage{algpseudocode}
\usepackage{pifont}
\usepackage[colorinlistoftodos]{todonotes}

\newtheorem{theorem}{Theorem}
\newtheorem{remark}{Remark} 
\newtheorem{proposition}{Proposition}

\newcommand{\beqn}{\begin{equation}}
\newcommand{\eeqn}{\end{equation}}
\newcommand{\beqa}{\begin{eqnarray}}
\newcommand{\eeqa}{\end{eqnarray}}
\newcommand{\beqas}{\begin{eqnarray*}}
\newcommand{\eeqas}{\end{eqnarray*}}

\DeclareMathOperator*{\argmax}{arg\,max}

\presetkeys{todonotes}{color=green!30}{}
\newcommand{\vz}{{{\bf z}} }
\newcommand{\vy}{ {{\bf y}} }
\newcommand{\vx}{ {{\bf x}} }
\newcommand{\mh}{ {{\bf H}} }

\begin{document}

\title{Interference Cancellation and Iterative Detection for Orthogonal Time Frequency Space Modulation}
\author{P. Raviteja, Khoa T. Phan,  Yi Hong, and Emanuele Viterbo
\thanks{The authors are with ECSE Department, Monash University, Clayton, VIC 3800, Australia.
Email: \{raviteja.patchava, khoa.phan, yi.hong, emanuele.viterbo\}@monash.edu.}}
\maketitle

\begin{abstract}
The recently proposed orthogonal time frequency space (OTFS) modulation technique was shown to provide significant error performance advantages over orthogonal frequency division multiplexing (OFDM) in
Doppler channels. In this paper, we first derive the explicit input--output  relation describing OTFS modulation and demodulation (mod/demod) for delay--Doppler channels. We then analyze the cases of ({\em i}) ideal pulse-shaping waveforms that satisfy the bi-orthogonality conditions, and ({\em ii}) rectangular waveforms which do not.  We show that while only {\em inter-Doppler interference (IDI)} is present in the first case, additional {\em inter-carrier interference (ICI)} and {\em inter-symbol interference (ISI)} occur in the second case. We next analyze the interferences and develop a novel low-complexity yet efficient message passing (MP) algorithm for joint interference cancellation (IC) and symbol detection.  
While  ICI and ISI are eliminated through appropriate phase shifting, IDI can be mitigated by adapting the MP algorithm to account for only the largest interference terms. The proposed MP algorithm can effectively compensate for a wide range of channel Doppler spreads.
Our results indicate that OTFS using practical rectangular waveforms can achieve the performance of OTFS using ideal but {\em non-realizable} pulse-shaping waveforms. Finally, simulations results demonstrate the superior error performance gains of the proposed {\em uncoded} OTFS schemes over OFDM under various channel conditions. 
\end{abstract}
\begin{IEEEkeywords} 
Delay--Doppler channel, OTFS, message passing, time--frequency modulation.  
\end{IEEEkeywords}
\section{Introduction} 
Fifth-generation (5G) mobile systems are expected to accommodate an enormous number of emerging wireless applications with high data rate requirements such as real-time video streaming, and online gaming, connected and autonomous vehicles. While the orthogonal frequency division multiplexing (OFDM) modulation scheme currently deployed in fourth-generation (4G) mobile systems can achieve high spectral efficiency for time-invariant frequency selective channels, it is not robust to time-varying channels with high Doppler spread (e.g., high-speed railway mobile communications). Hence, new modulation techniques that are robust to channel time-variations have been extensively explored. 

To cope with time-varying channels, one existing approach is to shorten the OFDM symbol duration so that the channel variations over each symbol appear inconsequential \cite{proakis}. However, one major drawback is the reduced spectral efficiency due to cyclic prefix
(CP). Another approach exploits time--frequency signaling \cite{Matz}, \cite{Liu}. In \cite{Dean} the authors have introduced the frequency-division
multiplexing with frequency-domain cyclic prefix (FDM-FDCP), which 
can efficiently compensate for channel Doppler spread. In high-Doppler low-delay spread channels, FDM-FDCP is shown to outperform OFDM at the same spectral efficiency. The performance of FDM-FDCP under other channel conditions are yet to be studied. 

A new time--frequency modulation technique called orthogonal time frequency space
(OTFS) was recently proposed in \cite{Hadani,otfs_white}, which shows significant advantages over OFDM in delay--Doppler channels. The delay--Doppler domain provides an alternative representation of a time-varying channel geometry modeling mobile terminals and reflectors \cite{matz_book,Jakes}. Leveraging on this representation, the OTFS modulator spreads each information symbol over a two dimensional (2D) orthogonal basis function, which spans across the entire time--frequency domain required to transmit a frame. The set of basis functions is specifically designed to combat the dynamics of the
time-varying multi-path channels. In \cite{Hadani}, a general framework of OTFS based on ideal pulse-shaping waveforms was introduced. A {\it coded} OTFS system with  turbo equalization was compared with coded OFDM, showing remarkable gains. In \cite{Hadani1}, since mm-wave channels incur  high frequency dispersion, OTFS is shown to outperform OFDM significantly.

In this paper, we analyze the input--output relation describing {\em uncoded} OTFS mod/demod for delay--Doppler channels using general pulse-shaping waveforms. The relation reveals the effects of the inverse symplectic finite Fourier transform (ISFFT) and SFFT operations interpreted as pre- and post-processing blocks applied to a time--frequency signaling scheme. We then analyze the cases of ({\em i}) ideal pulse-shaping waveforms that satisfy the bi-orthogonality conditions, and ({\em ii}) practical rectangular waveforms which do not.
Unlike previous works \cite{NewOTFS}, \cite{NewOTFS1}, we assume {\it no} CP in the second case. We show that, while only  {\it inter-Doppler interference} (IDI) is present in the ideal waveform case due to unavoidable fractional Doppler effects, additional {\it inter-carrier interference} (ICI) and {\it inter-symbol interference} (ISI) occur in the latter case due to imperfect {\it bi-orthogonality} in time--frequency domain of the rectangular waveforms.  

The delay--Doppler channel model with a small number of paths, with varying delay and Doppler values, provides a sparse representation of the communication channel. %we next explicitly characterize the interferences paving the way for the development of interference cancellation (IC) techniques. 
We then propose a low-complexity message passing (MP) algorithm for a joint interference cancellation (IC) and detection, which takes advantage of the inherent delay--Doppler channel sparsity. 
The MP algorithm is based on a sparse factor graph and uses Gaussian approximation of the interference terms to further reduce the complexity. The approach is similar to \cite{mp_ga}, where it was applied to massive MIMO without the advantage of channel sparsity. The complexity and convergence of the MP algorithm are analyzed. 
In the MP algorithm, while the ICI and ISI can be eliminated by suitable phase shifting, the IDI can be mitigated by adapting the MP algorithm to account for only the largest interference terms.  Consequently, the proposed MP algorithm can effectively compensate for a wide range of channel Doppler spreads. Further, our results show that OTFS using practical rectangular waveforms can achieve the performance of OTFS using ideal but {\em non-realizable} pulse-shaping waveforms.  
Simulations results illustrate the superior performance gains of the proposed {\em uncoded} OTFS schemes over OFDM under various channel conditions. 

The rest of the manuscript is organized as follows. Section II recalls the OTFS mod/demod and derives the corresponding input--output relation.   In Section III, we analyze the time--frequency domain and delay--Doppler domain relations for the ideal waveform case. Section IV is dedicated to the case of OTFS using rectangular waveforms. Section V proposes the MP algorithm for the joint IC and detection. Simulation results are presented in Section VI followed by the conclusions in Section VII.   The proofs are relegated to Appendix at the end of the manuscript. 
\begin{figure*}
\centering
\includegraphics[width=6in]{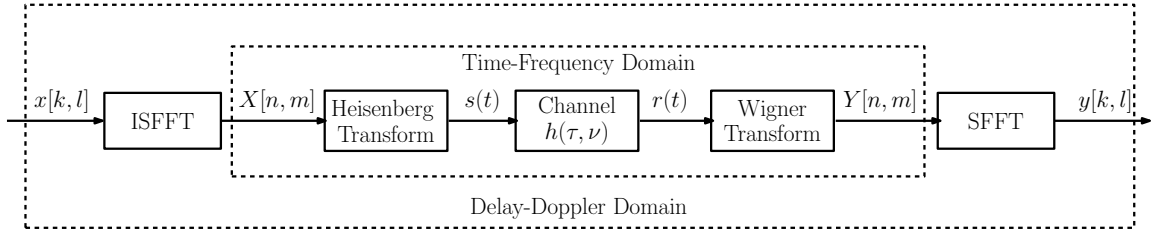}
\caption{OTFS mod/demod}
\label{sys_fig}
\end{figure*}
\section{System model} 
In this section, we first recall the basic concepts in OTFS and then present the explicit analysis of OTFS mod/demod.  More importantly, we derive the input--output relation of OTFS mod/demod for delay--Doppler channels. 
\subsection{Basic OTFS concepts/notations} 
We follow the notations in \cite{Hadani,Ravi_WCNC} summarized below:

-- The {\em time--frequency signal plane} is discretized to a grid by sampling time and frequency axes at intervals $T$ (seconds) and $\Delta f$ (Hz), respectively, i.e., 
\beqn 
\Lambda = \bigl\{(nT,m\Delta f),\; n=0,\hdots,N-1, m=0,\hdots,M-1\bigr\} \nonumber 
\eeqn 
for some integers $N, M >0$.

%-- A packet burst has duration $NT$ and bandwidth $M\Delta f$. 

-- Modulated {\em time--frequency samples} $X[n,m], n=0,\hdots,N-1, m=0,\hdots,M-1$ are transmitted over an OTFS frame with duration $T_f = NT$ and occupy a bandwidth $B = M\Delta f$. 

-- Transmit and receive pulses (or waveforms) are denoted by $g_{\text{tx}}(t)$ and $g_{\text{rx}}(t)$. Let $A_{g_{rx},g_{tx}}(t,f)$ denote the {\em cross-ambiguity function} between  $g_{\text{tx}}(t)$ and $g_{\text{rx}}(t)$, i.e., 
\beqn 
A_{g_{\text{rx}}, g_{\text{tx}}}(t,f)\triangleq \int g_{\text{rx}}^*(t'-t) g_{\text{tx}}(t') e^{-j2\pi f(t' -t)} dt'. \label{cross}
\eeqn 

-- The delay--Doppler plane is discretized to an information grid: 
\beqn 
\Gamma = \Bigl\{\left(\frac{k}{NT},\frac{l}{M\Delta f}\right),\; k=0,\hdots,N-1, l=0,\hdots,M-1\Bigr\}, \nonumber 
\eeqn 
where $1/M\Delta f$ and $1/NT$  represent the quantization steps of the delay and Doppler frequency, respectively\footnote{Note that the first and second indexes, $k$ and $l$, in $\Gamma$ represent the Doppler and delay axis, respectively.}.

\begin{remark}{\em Parameter choice for OTFS systems.} 
Given a communications system with total bandwidth $B=M\Delta f$ and latency $T_f=NT$ constraints, we may choose $N$, $M$, $T$, (since $\Delta f=1/T$) over a time-varying channel with maximum delay $\tau_{\rm max}$ and maximum Doppler $\nu_{\rm max}$, among all channel paths.  
We can see that $T$ and $\Delta f$ determine the maximum supportable Doppler (i.e., $1/T$) and delay (i.e., $1/\Delta f$).
Hence, it is required that $\nu_{\rm max} < 1/T$ and $\tau_{\rm max} < 1/\Delta f$ so that $N$ and $M$ are determined. %(preferably much smaller $\ll$).  
To support a fixed data rate of $NM$ symbols per frame, depending on the channel conditions, we can choose a larger $T$ and smaller $\Delta f$, which results in a smaller $N$ and larger $M$, respectively, or vice versa. 

\end{remark}

\subsection{General OTFS mod/demod block diagram} 
The OTFS system diagram is given in Fig. \ref{sys_fig}. OTFS modulation is produced by a cascade of a pair of 2D transforms at both transmitter and receiver.  The modulator first maps the information symbols $x[k,l]$ in the delay--Doppler domain to samples $X[n,m]$  in the time--frequency domain using the {\em inverse symplectic finite Fourier transform} (ISFFT). Next, the {\em Heisenberg transform} is applied to $X[n,m]$ to create the time domain signal $s(t)$ transmitted over the wireless channel.
At the receiver, the time-domain signal $r(t)$ is mapped to the time--frequency domain through the {\em Wigner transform} (the inverse of the Heisenberg transform), and then to the delay--Doppler domain using SFFT for symbol demodulation.  

\subsection{OTFS modulation} 
Consider a set of $NM$ information symbols $\{x[k,l], k=0,\ldots,N-1, l=0,\ldots, M-1\}$ from a modulation alphabet of size $Q$ $\mathbb{A} = \{ a_1, \cdots, a_{Q} \}$ (e.g. QAM symbols), %\todo{\tiny{These are elements of a set, so they should be denoted $a_1,\hdots,a_{Q}$}} 
which are arranged on the delay--Doppler grid $\Gamma$.
 
The OTFS transmitter first maps symbols $x[k,l]$ to $NM$ samples $X[n,m]$ on the time--frequency grid $\Lambda$ using the ISFFT as follows: 
\beqn
X[n,m] = \frac{1}{\sqrt{NM}}\sum_{k=0}^{N-1}\sum_{l=0}^{M-1} x[k,l] e^{j2\pi\bigl(\frac{nk}{N}-\frac{ml}{M}\bigr)} \label{iSFFT}
\eeqn 
for $n=0,\hdots,N-1, m=0,\hdots,M-1$. 
% We have assumed rectangle transmit windowing function
% that multiplies the modulation symbols in the time--frequency domain.

Next, a time--frequency modulator converts the samples $X[n,m]$   to a continuous time  waveform $s(t)$ using a transmit  waveform 
$g_{\text{tx}}(t)$ as
\beqn
s(t) = \sum_{n=0}^{N-1}\sum_{m = 0}^{M-1}   X[n,m] g_{\text{tx}}(t-nT) e^{j2\pi m \Delta f(t-nT)}. \label{otfs1}
\eeqn 
As noted in \cite{Hadani},  (\ref{otfs1}) is  also referred to in the mathematical literature as the (discrete) Heisenberg transform  \cite{Heisenberg}, parametrized by $g_{\text{tx}}(t)$.
% \begin{remark}  
% \textcolor{blue}{
% The operation  (\ref{otfs1}) generalizes OFDM which only maps information symbols from frequency domain to time domain. 
% If $g_{\text{tx}}(t)$ is a rectangular pulse waveform of duration $T$, then (\ref{otfs1}) reduces to the conventional inverse discrete Fourier transform. When $N=1$, the inner box in Fig. \ref{sys_fig} is simply an OFDM system with $M$ subcarriers. Therefore, one OTFS symbol (or frame) can be viewed as a ISFFT pre-coding applied on $N$ consecutive independent OFDM symbols. The case of rectangular waveforms will be studied in detail in Section IV. }
% \end{remark}  
% \todo{\tiny @Khoa: Since we do not want to emphasize too much on the relationship between OTFS and OFDM  (as we plan to treat ISSF and SFFT as pre-and post-coding OFDM separately in the future), should we remove (or just mention very briefly) Remarks 2, 3, and so on? (especially the highlighted text which is exactly what we want to do)
%}
\subsection{Wireless transmission and reception} 
The signal $s(t)$ is transmitted over a time-varying channel with complex baseband channel impulse response $h(\tau,\nu)$, 
which characterizes the channel response to an impulse with delay $\tau$ and Doppler $\nu$ \cite{Jakes}. The received signal $r(t)$ is given by (disregarding the noise to simplify notation):
\beqn
r(t) = \int \int h(\tau,\nu) s(t-\tau) e^{j2\pi\nu(t-\tau)}d\tau d\nu. \label{delay-doppler}
\eeqn
Equation (\ref{delay-doppler}) represents a continuous Heisenberg transform  parametrized by $s(t)$ \cite{Hadani}. 
Since typically there are only a small number
of reflectors in the channel with associated delays and Dopplers, very few
parameters are needed to model the channel in the delay--Doppler domain. 
The  sparse representation of the channel $h(\tau,\nu)$  is given as: 
\begin{equation}
h(\tau,\nu) = \sum_{i=1}^{P} h_i \delta(\tau-\tau_i) \delta(\nu-\nu_i)
\label{del_dop}
\end{equation}
where $P$ is the number of propagation paths, $h_i$, $\tau_i$, and $\nu_i$ represent the path gain, delay, and Doppler shift (or frequency) associated with $i$-th path, respectively, and $\delta(\cdot)$ denotes the Dirac delta function. 
%Here, we assume that all paths have different delays.
%\todo{\tiny Why do we need this?--I think not required, this condition simplifies only the channel estimation phase} 
We denote the delay and Doppler taps for $i$-th path as follows:
\beqn
\tau_i = \frac{l_{\tau_i}}{M\Delta f},\;\;\nu_i = \frac{k_{\nu_i} + \kappa_{\nu_i}}{NT}
\label{delaytap}
\eeqn 
for integers $l_{\tau_i}, k_{\nu_i}$ and real $-\frac{1}{2}< \kappa_{\nu_i} \leq \frac{1}{2}$.
%As mentioned, it is assumed that $\tau_{\rm max} < 1/\Delta f$ and $\nu_{\rm max} < 1/T$.
Specifically, $l_{\tau_i}$ and $k_{\nu_i}$ represent the indexes of the delay tap and Doppler  tap,  corresponding to (continuous) delay $\tau_i$ and Doppler frequency $\nu_i$, respectively. 
 We will refer to $\kappa_{\nu_i}$ as the {\em fractional} Doppler since it represents the fractional shift from the nearest Doppler tap $k_{\nu_i}$. We do not need to consider fractional delays since the resolution of the sampling time $\frac{1}{M\Delta f}$ is sufficient to approximate the path delays to the nearest sampling points in typical wide-band systems \cite{wc_book}. 
%\todo{\tiny say why only care of fractional Doppler and not fractional delay: Is it because the fractional delay does not %produce as much interference as fractional Doppler? } 

\subsection{OTFS demodulation}
At the receiver, a matched filter computes the cross-ambiguity function $A_{g_{\text{rx}},r}(t,f)$: 
\beqn
Y(t,f) = A_{g_{\text{rx}},r}(t,f) \triangleq \int g_{\text{rx}}^*(t'-t) r(t') e^{-j 2 \pi f (t'-t)} dt'. \label{wigner1}
\eeqn
The matched filter output  is obtained by sampling $Y(t,f)$ as
\beqn
Y[n,m] = Y(t,f)|_{t = nT, f = m \Delta f}. \label{wigner}
\eeqn 
for $n=0,\hdots,N-1$ and  $m=0,\hdots,M-1$. 
Operations (\ref{wigner1}) and (\ref{wigner}) are referred as the  {\em Wigner transform}.
% The relationship between the 
% \begin{lemma}
% \label{le1}
% The value of $Y(t,f)$ can be expressed as 
% \begin{equation}
% Y(t,f) = h(t,f) 
% \end{equation}
% \end{lemma}
% {\em Proof:}  The proof is relegated to the Appendix. \hfill $\blacksquare$\\
% \begin{remark}
% \textcolor{blue}{
% The Wigner transform is a generalization of the OFDM receiver, which maps the received time domain signal to the frequency domain modulated symbols. 
% When $g_{\text{rx}}(t)$ is a rectangular pulse, it corresponds to the discrete Fourier transform in  OFDM. }
% \end{remark
%}
In the following theorem, we characterize the relationship between time--frequency output samples $Y[n,m]$ and input samples  $X[n,m]$.
\begin{theorem} \label{Theorem1}
{\it OTFS time--frequency domain analysis.} The following input--output relation of OTFS in time--frequency domain  is given by:
\begin{equation}
Y[n,m] = \sum_{n'=0}^{N-1} \sum_{m'=0}^{M-1} H_{n,m}[n',m'] X[n',m'],
\label{eq_freq_time}
\end{equation}
where 
\begin{align}
 H_{n,m}[n',m'] & = \int\int h(\tau,\nu) A_{g_{rx},g_{tx}}((n-n')T-\tau,(m-m')\Delta f-\nu) \nonumber \\
& \qquad \quad e^{j 2\pi (\nu+m'\Delta f) ((n-n')T-\tau)} e^{j 2\pi \nu n' T}d\tau d\nu.
\label{eq_h_freq}
\end{align}
\label{thr_freq_time}
\end{theorem}
{\em Proof:}  The proof is given in Appendix \ref{app_thr_freq_time}. \hfill $\blacksquare$\\
We can see that the terms $H_{n,m}[n',m']$ include the combined effects of the transmit pulse, channel, and receive pulse. 
Note that results similar to this theorem have been presented for the case of pulse-shaped (PS) OFDM \cite{durisi,molisch}.

Next, the SFFT is applied on the samples $Y[n,m]$ to obtain symbols $y[k,l]$ in the delay--Doppler domain:
\begin{align}
y[k,l] & = \frac{1}{\sqrt{NM}} \sum_{n=0}^{N-1} \sum_{m=0}^{M-1} Y[n,m] e^{-j2\pi\bigl(\frac{nk}{N}-\frac{ml}{M}\bigr)}.
\label{rx_sfft}
\end{align}
%\todo{\tiny{Should be without $\frac{1}{NM}$? please check}}

Theorem \ref{Theorem1}  provides the basis of the study of OTFS in two special cases, namely using {\em ideal waveforms} (Section III) and more practical {\em rectangular waveforms} (Section IV).  We will obtain explicit input-output relations using the delay--Doppler channel model (\ref{del_dop}) for both cases.

%%%%%%%%%%%%%%%%%%%%%%%%%%%%%%%%%%%%%%%%%%%%%%%%%%%%%%%
\section{OTFS with ideal waveforms}
The $g_{\text{rx}}(t)$ and $g_{\text{tx}}(t)$ pulses are said to be ideal if they satisfy the  {\em bi-orthogonal property} \cite{Hadani}
%where $n\in[1,N-1]$, and $m\in[1,M-1]$ and $A_{g_{\text{rx}}, g_{\text{tx}}}(t,f) = 1$ for $t \in (-\tau_{\rm max},\tau_{\rm max})$, $f \in (-\nu_{\rm max},\nu_{\rm max})$. 
%
%\todo{\tiny is this notation in (12) correct ?-- I think $q_{\tau_{\rm max}}(t) q_{\nu_{\rm max}}(t)$ is not required }
\begin{align}
A_{g_{\text{rx}}, g_{\text{tx}}}(t,f)|_{t = nT + (-\tau_{\rm max},\tau_{\rm max}), f = m \Delta f + (-\nu_{\rm max},\nu_{\rm max})} = \delta[n]\delta[m] q_{\tau_{\rm max}}(t) q_{\nu_{\rm max}}(f)
\label{bi-ortho}
\end{align} 
where $q_a(x)=1$ for $x\in (-a,a)$ and zero otherwise.
Equivalently, $A_{g_{\text{rx}}, g_{\text{tx}}}(t,f) = 0$ for $t \in (nT-\tau_{\rm max},nT+\tau_{\rm max})$ and $f \in (m\Delta f-\nu_{\rm max},m\Delta f+\nu_{\rm max})$, for all values of $n,m$ except for $n=0,m=0$, where $A_{g_{\text{rx}}, g_{\text{tx}}}(t,f) = 1$ for $t \in (-\tau_{\rm max},\tau_{\rm max})$ and $f \in (-\nu_{\rm max},\nu_{\rm max})$. 

Unfortunately, ideal pulses cannot be realized in practice but can be approximated by waveforms with a support concentrated as much as possible in time and in frequency, given the constraint imposed by the uncertainty principle. Nevertheless, it is important to study the error performance of OTFS with ideal waveforms since it serves as a lower bound on the performance of OTFS with practically realizable waveforms such as rectangular waveforms, etc.

%todo{Dear Ravi, can you elaborate on the negative delay and Doppler and how they are mapped to the delay--Doppler plane? So far, it seems that the %grid is for positive delay and Doppler only. Do we need to modify the definitions of the grid somehow? }
%\todo{\tiny{Is it $nT + (-\tau_{\rm max},\tau_{\rm max})$ or $nT \in (-\tau_{\rm max},\tau_{\rm max})$?} It is '+' only as it %can cover any value in the entire interval depends on $\tau_i$ and $\nu_i$ }
% \textcolor{blue}{Is the above explanation equivalent to the following:
% $$
% A_{g_{\text{rx}}, g_{\text{tx}}}(t,f)|_{t = nT \in (-\tau_{\rm max},\tau_{\rm max}), f = m \Delta f \in (-\nu_{\rm max},\nu_{\rm max})} = \delta(n)\delta(m)
% $$
% Note that we have defined $A_{g_{\text{rx}}, g_{\text{tx}}}(t,f)$ in (\ref{cross}). If they are, I think we should use the latter one for ideal pulse definition}
% \todo{\tiny{see comment}}

\subsection{Time--frequency domain analysis} 
For ideal waveforms, the following result was given in \cite{Hadani} without proof. Here, we show that it can be obtained as a special case of Theorem $1$. 
{\proposition{For ideal pulses,  the following result can be obtained:    
\beqn \label{eq:receivedsig_Y}
Y[n,m] = H_{n,m}[n,m]X[n,m] 
\eeqn 
%\todo{Should it be $H_{n,m}[n,m]$?}
where
\beqn
H_{n,m}[n,m] = \int \int h(\tau,\nu) {e^{j 2\pi \nu nT}} e^{-j 2\pi(\nu + m\Delta f)\tau} d\tau d\nu. \nonumber
\eeqn 
} \label{ideal_case}}
%\todo{Should it be $e^{j 2\pi \nu nT}$?}
{\em Proof:} From (\ref{eq_h_freq}), we observe that the value of $H_{n,m}[n',m']$ is non-zero only at $n' = n$ and $m' = m$ for the ideal pulses satisfying the bi-orthogonal property (\ref{bi-ortho}). Hence, the result in (\ref{eq:receivedsig_Y}) follows from (\ref{eq_freq_time}) by considering only the term with $n'=n$, $m'=m$ in the summations. 
%The proof is presented in the Appendix. 
\hfill $\blacksquare$ \\
\subsection{Delay--Doppler domain analysis}
\subsubsection{Input--output relationship}
We now apply SFFT on $Y[n,m]$ in (\ref{eq:receivedsig_Y}) to obtain the symbols $y[k,l]$ in the delay--Doppler domain. The following proposition, given in \cite{Hadani} without proof, describes the  input--output  relation in delay--Doppler domain. 
{\proposition\label{pro2}{For ideal pulses, the following input-output relation holds:  
\begin{align}
y[k,l] & =  \frac{1}{NM}\! \sum_{k'=0}^{N-1} \sum_{l'=0}^{M-1} x[k',l'] h_w[k-k', l-l'],
\label{rx_symbol}
\end{align}
where  $h_w[\cdot,\cdot]$ is a sampled version of the impulse response function: 
\beqn
h_w[k-k', l-l'] = h_w(\nu,\tau)|_{\nu=\frac{k-k'}{NT},\tau=\frac{l-l'}{M\Delta f}} 
\label{h_w}
\eeqn 
for $h_w(\nu,\tau)$ being the circular convolution of the channel response with the SFFT of a rectangular windowing function in the time-frequency domain: 
\begin{align}
h_w(\nu,\tau) &= \int \int h(\tau',\nu') w(\nu - \nu', \tau -\tau') e^{-j2\pi\nu\tau}d\tau' d\nu', \label{h_w_time}\\
w(\nu,\tau) &= \sum_{n=0}^{N-1} \sum_{m=0}^{M-1} 1\cdot e^{-j 2\pi (\nu nT - \tau m\Delta f)}. \label{window}
\end{align}
}}
{\em Proof:}  The proof is relegated to the Appendix \ref{app_pro2}. \hfill $\blacksquare$ \\  

\subsubsection{Inter-Doppler interference (IDI) analysis} From (\ref{rx_symbol}), we can see that a received signal $y[k,l]$ is a linear combination of all the transmitted signals $x[k',l'], k'=0,\hdots,N-1, l'=0,\hdots,M-1$. Consequently, the input-output relation (\ref{rx_symbol}) can be represented as a  linear system with  $NM$  variables $x[k',l']$. Since $N$ and $M$ tend to be very large for practical OTFS systems, the detection complexity can be prohibitive.  In the following, by using (\ref{del_dop}) as the sparse representation of the delay--Doppler channel,  (\ref{rx_symbol}) reduces to a sparse linear system,  where each received signal can be approximately expressed as a linear combination of only a few transmitted signals. Such sparsity will then be exploited in Section V to devise a low-complexity yet efficient iterative detection algorithm based on message passing on the factor graph representation.

By substituting (\ref{del_dop}) and (\ref{window}) into (\ref{h_w_time}), we obtain:
\begin{align}
h_w(\nu,\tau)  &= \sum_{i=1}^{P} h_i e^{-j2\pi\nu_i\tau_i} \, w(\nu - \nu_i, \tau -\tau_i)
\nonumber \\ 
& = \sum_{i=1}^{P} \, h_i e^{-j2\pi\nu_i\tau_i} \, \mathcal {G} (\nu,\nu_i) \, \mathcal {F} (\tau,\tau_i),
\end{align}%\todo[inline]{\tiny remove the second line in (8) if we run out of space. Swap F and G}
where we have denoted: 
\beqa 
\mathcal {F} (\tau,\tau_i) &\triangleq & \sum_{m'=0}^{M-1} e^{j 2\pi (\tau -\tau_i) m'\Delta f},\nonumber\\
\mathcal {G} (\nu,\nu_i) &\triangleq& \sum_{n'=0}^{N-1} e^{-j 2\pi (\nu - \nu_i) n'T}.\nonumber
\eeqa 
% \textcolor{blue}{For clarity, we evaluate separately $h_w(0,0)$ and $h_w\Bigl(\frac{k-n}{NT}, \frac{l-m}{M\Delta f}\Bigr)$ for $n \ne k$ and $m \ne l$.}
Let us first evaluate $\mathcal {F} (\tau,\tau_i)$ at $\tau = \frac{l-l'}{M\Delta f}$ as: 
\begin{align}
\mathcal {F} \, \left(\frac{l-l'}{M\Delta f},\tau_i\right) = \sum_{m'=0}^{M-1} e^{j \frac{2\pi}{M} (l - l' - l_{\tau_i}) m'}= \frac{e^{j {2\pi} (l - l' - l_{\tau_i}) }-1}{e^{j \frac{2\pi}{M} (l - l' - l_{\tau_i})}-1}
\label{f_eq}
\end{align}
recalling that $l_{\tau_i}$ is the  delay tap of $i$-th path with a delay $\tau_i$ defined in (\ref{delaytap}). 
From (\ref{f_eq}), we see that
\[
\mathcal {F} \, \left(\frac{l-l'}{M\Delta f},\tau_i\right) = \left.
 \begin{cases}
 M, &  [l-l'-l_{\tau_i}]_M = 0, \\
 0, & \mbox{otherwise, }
 \end{cases}
 \right.
\]
where $[\cdot]_M$ represents mod $M$ operation, i.e., it evaluates to $M$ for $l' = [l-l_{\tau_i}]_M $ and is zero otherwise.

Similarly, we can evaluate: 
\begin{align}
\mathcal {G} \, \left(\frac{k-k'}{NT},\nu_i \right) & = \frac{e^{j {2\pi} (k - k' - k_{\nu_i} - \kappa_{\nu_i}) }-1}{e^{j \frac{2\pi}{N} (k - k' - k_{\nu_i}- \kappa_{\nu_i})}-1} . 
\label{g_val}
\end{align}
Due to the fractional $\kappa_{\nu_i}$, we can see that for given $k$, $\mathcal {G} \, \left(\frac{k-k'}{NT},\nu_i \right) \ne 0,\forall n$.

We will show that the magnitude of $\frac{1}{N} \mathcal {G} \left(\frac{k-k'}{NT},\nu_i \right)$ has a peak at $k'= k - k_{\nu_i}$ and decreases as $k'$ moves away from $k - k_{\nu_i}$.
From (\ref{g_val}), after some manipulations, we have:   
  \begin{align}
 \left|\frac{1}{N} \mathcal {G} \left(\frac{k-k'}{NT},\nu_i \right)\right|  = \left|\frac{\sin (N\theta)}{N\sin (\theta)}\right| \label{sinc}
 \end{align}
 where we set $\theta \triangleq \frac{\pi}{N} \left( k-k'-k_{\nu_i} - \kappa_{\nu_i}\right)$. 
It can be easily shown that: 
 \begin{align}
 \left|\frac{\sin (N\theta)}{N\sin (\theta)}\right| & = \left|\frac{\sin((N-1)\theta)\cos(\theta) + \sin(\theta)\cos((N-1)\theta) }{N\sin(\theta)}\right|
 \nonumber \\ 
 & \le \frac{N-1}{N}\left|\cos(\theta)\right|+\frac{1}{N}.
 \label{UpperBDSinOverSin}
 \end{align} 
Here, we used the inequality, $|\sin(N\theta)| \leq N|\sin(\theta)|$, which can be proven by induction. The upper bound (\ref{UpperBDSinOverSin}) is tight for small values of $\theta$ (when both sides are close to $1$) and it has a peak at the smallest value of $\theta$ when $k'=k-k_{\nu_i}$. As $|\theta|$ increases (due to $k'$ moving away from $k-k_{\nu_i}$), the upper bound decreases with (approximate) slope  of $\frac{\pi}{N} \left( k-k'-k_{\nu_i} - \kappa_{\nu_i}\right)$. 
%To summarize, we can see that $\left|\frac{1}{N} \mathcal {G} \left(\frac{k-k'}{NT},\nu_i \right)\right|$ has a peak at $k' = k - k_{\nu_i}$ and decreases as $k'$ moves away from $k-k_{\nu_i}$. 
Since $N$ is quite large in OTFS, the function decreases rapidly.   

% Similarly, $\left|\frac{1}{N} \mathcal {G} \left(\frac{k-k'}{NT},\nu_i \right)\right|$ decreases with increasing $k'=k - k_{\nu_i}, k - k_{\nu_i}+1, k - k_{\nu_i}+2,\hdots$.

From the above analysis, we need to consider only a small number $2N_i +1$, for some $N_i>0$, of significant values of $\mathcal {G} \left(\frac{k-k'}{NT},\nu_i \right)$ in (\ref{g_val}) around the peak $k - k_{\nu_i}$, i.e., $[k - k_{\nu_i} - N_i]_N \le k' \le [k - k_{\nu_i} + N_i]_N$, where $N_i\ll N$. Using this approximation, we can now express the receive signal $y[k,l]$ in (\ref{rx_symbol}) as: %\todo{\tiny need too say the approx error is within $1/N$ }
\begin{align}
%y(k,l) & = \frac{1}{NM} \sum_{n=0}^{N-1} \sum_{m=0}^{M-1}  x[n,m] \\
%& \sum_{i=1}^{P} \, a'_i \, \mathcal {F} \left(\frac{l-m}{M\Delta f},\tau_i\right) \, \mathcal {G} \left(\frac{k-n}{NT},\nu_i\right) \nonumber \\
y[k,l] {\approx} & \sum_{i=1}^{P} \sum_{k'=[k-k_{\nu_i}-N_i]_N}^{[k-k_{\nu_i}+N_i]_N}   \left(\frac{e^{j {2\pi} (k-k_{\nu_i}- k' - \kappa_{\nu_i}) }-1}{N e^{j \frac{2\pi}{N} (k-k_{\nu_i}- k' - \kappa_{\nu_i})}-N}\right)  h_i e^{-j2\pi\nu_i\tau_i} \, x\left[k', [l - l_{\tau_i}]_M\right] \nonumber \\
 {\color{black}\approx} & \sum_{i=1}^{P} \sum_{q=-N_i}^{N_i}   \left(\frac{e^{j {2\pi} (-q - \kappa_{\nu_i}) }-1}{N e^{j \frac{2\pi}{N} (- q - \kappa_{\nu_i})}-N}\right) h_i e^{-j2\pi\nu_i\tau_i}  x\left[[k-k_{\nu_i}+q]_N, [l - l_{\tau_i}]_M\right].
\label{conv_eq}
\end{align}
In the simulation result section, we will demonstrate that for $N = 128$, by choosing $N_i=10$, no performance loss is incurred.
From (\ref{conv_eq}), we can see that the received signal $y[k,l]$ is a linear combination of $S = \sum_{i=1}^P 2N_i+1$ transmitted signals. Out of $2N_i+1$ transmitted signals in $i$-th path, the signal corresponding to $q=0$, $x\left[[k-k_{\nu_i}]_N, [l - l_{\tau_i}]_M\right]$, contributes the most and all the other $2N_i$ signals can be seen as  interferences. Such interferences are due to the transmitted signals that are neighbor to $x\left[[k-k_{\nu_i}]_N, [l - l_{\tau_i}]_M\right]$ in the Doppler domain and we refer to this interference as {\em inter-Doppler interference (IDI)}.
Further, the number of transmitted signals $S$ affecting the received signal $y[k,l]$ in (\ref{conv_eq}) is much smaller than $NM$ in (\ref{rx_symbol}). Hence, the graph (or linear system) describing (\ref{conv_eq}) is sparsely-connected. 

% From (\ref{conv_eq}), we can see that with the fractional Doppler, the transmitted signal \textcolor{blue}{not only shifts by the delay and Doppler taps but also affects the neighboring Doppler taps} \todo{English usage? Consider revising for better clarity}. We refer to this interference on the neighboring Doppler taps as {\em inter-Doppler interference (IDI)}. \textcolor{blue}{We can see that in (\ref{conv_eq}), each received signal $y[k,l]$ is a linear combination of only $\sum_{i=1}^P 2N_i+1$ transmitted signals, which is much smaller than $NM$ in (\ref{rx_symbol}). Hence, the graph (or linear system) describing (\ref{conv_eq}) is sparsely-connected.} %\todo{\tiny{A better explanation why so-called IDI may be needed here.}}
\subsubsection{Special channel model cases}
\label{sec_special}
The above input-output expression simplifies for the following special cases. 

{\em i) Ideal channel}: Assuming $h(\tau,\nu) = \delta(\tau) \delta(\nu)$, the received signal becomes
\begin{align}
y[k,l] & = x[k,l]
\nonumber
\end{align}
and behaves as an AWGN channel as expected. 

{\em ii) No fractional Doppler} (i.e., $\kappa_{\nu_i}=0,\forall i$): Assuming that Doppler frequencies exactly coincide with Doppler taps, the received signal can be obtained by replacing $N_i = 0$ in (\ref{conv_eq}), i.e.,%\todo{\tiny Need to say why is this an approximation? }
\begin{align}
y[k,l] {\color{black}=} \sum_{i=1}^{P} h_i e^{-j2\pi\nu_i\tau_i} x[[k - k_{\nu_i}]_N, [l - l_{\tau_i}]_M].
\nonumber
\end{align}
%\todo{\tiny Should be $h'_i$ not $a'_i$. please check.}
For each path, the transmitted signal is circularly shifted by the delay and Doppler taps and scaled by the associated channel gain.

\section{OTFS with rectangular waveforms} 
Since the ideal pulses cannot be realized in practice, we now analyze the OTFS with the rectangular pulses at both the transmitter and receiver. These pulses do not satisfy the bi-orthogonality conditions and generate some interference which degrades the system performance. Here, we analyze the effect of such interference and show that it can be compensated to achieve the ideal pulses performance. 

We assume the rectangular pulse has amplitude $1/\sqrt{T}$ for $t \in [0,T]$ and $0$ at all other values, to have unit energy.

\subsection{Time--frequency domain analysis}
% Let us recall the time--frequency relation for the general case from Theorem \ref{thr_freq_time} and 
% \begin{equation}
% Y[n,m] = \sum_{n'=0}^{N-1} \sum_{m'=0}^{M-1} H_{n,m}[n',m'] X[n',m'], \label{time-frequency-relation}
% %\label{eq_freq_time}
% \end{equation}
% where 
% \begin{align*}
% & H_{n,m}[n',m'] = \nonumber \\
% & \int\int h(\tau,\nu) A_{g_{rx},g_{tx}}((n-n')T-\tau,(m-m')\Delta f-\nu) \nonumber \\
% & \qquad \quad e^{j 2\pi (\nu+m'\Delta f) ((n-n')T-\tau)} e^{j 2\pi \nu n' T}d\tau d\nu.
% \end{align*}
For the rectangular pulses, we can see that the cross-ambiguity term in the time--frequency relation of Theorem \ref{thr_freq_time}, $A_{g_{\text{rx}},g_{\text{tx}}}((n-n')T-\tau,(m-m')\Delta f-\nu)$ is non-zero for $|\tau|< \tau_{\max}$ $|\nu|< \nu_{\max}$  only when $n'=n$ and $n'=n-1$, since $g_{\text{tx}}$ and $g_{\text{rx}}$ are  pulses of duration $T$ and $\tau_{\rm max}\ll T$. Hence, the time--frequency relation (\ref{eq_freq_time}) becomes:
\begin{align}
& Y[n,m] = \sum_{n'=n-1}^{n} \sum_{m'=0}^{M-1} H_{n,m}[n',m'] X[n',m'] \nonumber\\
& = H_{n,m}[n,m] X[n,m] + \sum_{m'=0,m'\neq m}^{M-1} H_{n,m}[n,m'] X[n,m'] \nonumber \\
& \quad {} + \sum_{m'=0}^{M-1} H_{n,m}[n-1,m'] X[n-1,m'].
\label{eq_freq_time_rect}
\end{align}
The second term in (\ref{eq_freq_time_rect}) can be seen as the total interferences from the samples $X[n,m']$ at different frequencies $m' \ne m$ but same time slot $n$ as the current sample $X[n,m]$. On the other hand, the third term in (\ref{eq_freq_time_rect}) accumulates the interferences from the samples $X[n-1,m']$ in the previous time slot $n-1$. Hence, we call the second and third terms as the {\em inter carrier interference (ICI)} and {\em inter symbol interference (ISI)}, respectively.  
% The second and third terms in (\ref{eq_freq_time_rect}) represent the {\em inter carrier interference (ICI)} and {\em inter symbol interference (ISI)}, respectively. 
%\todo{Elaborate why inter carrier? why  inter symbol? since the readers may not be able to see/understand easily} 
The interferences depend on the delay $\tau$ and Doppler $\nu$ of the channel. In particular, they are affected by the value of the cross-ambiguity function $A_{g_{\text{rx}},g_{\text{tx}}}$ in $H_{n,m}[n',m']$. In the following, we focus on the cross-ambiguity function for ICI and ISI.

\subsubsection{ICI analysis} Fix $n, m$. 
We note that the cross-ambiguity function in the $H_{n,m}[n,m'], m' \ne m$ term of ICI, 
%$A_{\text{ici}}(m,m',\tau,\nu) = A_{g_{rx},g_{tx}}(-\tau,(m-m')\Delta f-\nu)$, 
$A_{g_{rx},g_{tx}}(-\tau,(m-m')\Delta f-\nu)$, is independent of $n$, and is computed for the $i$-th channel path with delay $\tau_i$ and Doppler $\nu_i$ (i.e., see (\ref{del_dop})) as:
\begin{align}
A_{\text{ici}} 
& \triangleq \int g_{\text{rx}}^*(t'+\tau_i) g_{\text{tx}}(t') e^{-j2\pi ((m-m')\Delta f-\nu_i)) (t'+\tau_i)} dt'.
\nonumber
\end{align}
%\todo{\tiny{$A_{\text{ici}} $ is dependent on $m, m' \ne m$. So we should associate $m, m'$ in $A_{\text{ici}} $.}}
We discard the dependency of $A_{\text{ici}} $ on $(m,m',\tau_i,\nu_i)$ for simplicity. Since the received  signal $r(t)$ is sampled at intervals of $T/M$ (or $1/(M\Delta f)$), we can compute $A_{\text{ici}}$ as:  
\begin{align}
A_{\text{ici}}  
& = \frac{1}{M}\sum_{p=0}^{M-1-l_{\tau_i}}  e^{-j2\pi((m-m')\Delta f-\nu_i)\left(\frac{p}{M\Delta f}+\tau_i\right)}.
\label{a_ici}
%& = \frac{1}{M} e^{-j2\pi((m-m')\Delta f-\nu)\tau} e^{-j2\pi((m-m')\Delta f-\nu)(M-l_{\tau_i})(T/M)}
\end{align}
%\todo{Is it $\frac{1}{T}$ or $\frac{1}{M}$? (the rectangular pulses have value $1/\sqrt{T}$). Check (26) too. --- it is $1/M$ only, $1/T$ will cancel with the $dt'=T/M$ term in integration.}
Recall that the pulses $g_{\text{tx}}$ and $g_{\text{rx}}$ have duration $T$, and $l_{\tau_i}$ is the delay tap defined in (\ref{delaytap}). 
%\todo{\tiny{Should it be $\frac{1}{T}$?  It is $1/M$ only as $T$ term gets canceled with rectangular waveform amplitude $\sqrt{1/T}$. %}}%Also, why $l_{\tau_i}$ for a particular path in the channel representation here? Are we using the model (\ref{del_dop})? Please check and mention explicitly.}
The amplitude of $A_{\text{ici}} $ is
\begin{align}
|A_{\text{ici}} | &= \frac{1}{M} \left| \sum_{p=0}^{M-1-l_{\tau_i}} e^{-j2\pi((m-m')\Delta f-\nu_i) \frac{p}{M\Delta f}} \right|  \nonumber \\ 
& = \frac{\left|e^{-j2\pi\left(m-m'-\frac{k_{\nu_i} + \kappa_{\nu_i}}{N}\right) \frac{M-l_{\tau_i}}{M}} -1\right|} { \left|M e^{-j2\pi\left(m-m'-\frac{k_{\nu_i} + \kappa_{\nu_i}}{N}\right) \frac{1}{M}} -M \right|}.
\nonumber
\end{align}
%\todo{It seems $+\tau_i$ in (25) is missing in the above expressions? Please check --- it is just a phase, doesn't impact the amplitude }
%\todo{\tiny{Better to show one previous manipulation step for easier understanding.}}
Similar to the analysis of (\ref{g_val}), we can observe that $|A_{\text{ici}} |$ decreases as $m'$ moves away from $m$. It implies that the ICI becomes less as the interfering subcarriers are further away from the interfered subcarrier. We can also see that an increase in Doppler (i.e., $k_{\nu_i} + \kappa_{\nu_i}$) increases the number of neighboring subcarriers that interfere with the present subcarrier. This is similar to the fractional Doppler effect studied for (\ref{g_val}). 
%\todo{Note easy to see. Please explain more.}
\subsubsection{ISI analysis}
Similar to the ICI analysis, the cross-ambiguity function in the $H_{n,m}[n-1,m']$ term of ISI,  $A_{\text{isi}} \triangleq A_{g_{rx},g_{tx}}(T-\tau,(m-m')\Delta f-\nu)$, is computed  for the $i$-th channel path as: 
\begin{align}
A_{\text{isi}} 
& = \frac{1}{M}\sum_{p=M-l_{\tau_i}}^{M-1}  e^{-j2\pi((m-m')\Delta f-\nu_i)\left(\frac{p}{M\Delta f}+\tau_i-T\right)}.
\label{a_isi}
\end{align}
The amplitude  $|A_{\text{isi}}|$ also has similar properties of $|A_{\text{ici}}|$, where it reduces as $m'$ moves away from $m$ implying that the ISI is smaller for interfering symbols further away (in the frequency axis) from the interfered symbol. %\todo{I corrected this statement. Please check}

Note that the terms that affect the ICI and ISI in the summations (\ref{a_ici}) and (\ref{a_isi}) are mutually exclusive, i.e., $p=0$ to $M-1-l_{\tau_i}$ contributes to ICI whereas $p=M-l_{\tau_i}$ to $M-1$ contributes to ISI. %\todo{\tiny{What do you mean? Please clarify}} 
This property helps in differentiating the ICI and ISI effects in delay--Doppler domain, which will be studied below.

\subsection{Delay--Doppler domain analysis}
We now characterize the input--output relation in delay--Doppler domain for OTFS with rectangular pulses. 
\begin{theorem}
\label{th_rect_dd}
The received signal $y[k,l]$ in delay--Doppler domain with the rectangular pulses can be written as
\begin{align}
% y[k,l]  = \sum_{i=1}^{P} & \sum_{k'=[k-k_{\nu_i}-N_i]_N}^{[k-k_{\nu_i}+N_i]_N}   h_i   
% \alpha_i(k,l,k') e^{j \theta_i(l)}  \nonumber\\
% & x\left[[k - k_{\nu_i} - k']_N, [l - l_{\tau_i}]_M\right]
\!\!\!\!\!\!\!\! \!\!\!\! \!\!\!\! y[k,l]  {\color{black}\approx} \sum_{i=1}^{P} & \sum_{q=-N_i}^{N_i}   h_i  e^{j2\pi \left(\frac{l-l_{\tau_i}}{M}\right) \left( \frac{k_{\nu_i}+\kappa_{\nu_i}}{N} \right)}  
\alpha_i(k,l,q) x\left[[k - k_{\nu_i} + q]_N, [l - l_{\tau_i}]_M\right]
\label{conv_eq_rect}
\end{align}
%\todo{\tiny{Change to $\alpha_i(k,l,k')$, $\theta_i(l)$, $\beta_i(k,k')$ etc.}}
where we have: 
\begin{align}
 \!\!\!\!\! \alpha_i(k,l,q) & =
  \begin{cases}
    \frac{1}{N} \beta_i(q)& \hspace{-2mm}   l_{\tau_i} \leq l < M\\
    \frac{1}{N}\left(\beta_i(q)-1\right) e^{-j2\pi \frac{[k - k_{\nu_i} + q]_N}{N}}  &   0\leq l<l_{\tau_i}
  \end{cases} \label{eq_alpha1} \\
\beta_i(q) & = \frac{e^{j {2\pi} (-q - \kappa_{\nu_i}) }-1}{e^{j \frac{2\pi}{N} (-q - \kappa_{\nu_i})}-1} \label{eq_beta1}.
% & \hspace{-17mm} e^{j\theta_i(l)}  =
% \begin{cases}
% e^{j\theta_{i,\text{ici}}(l)} = e^{j2\pi \frac{l}{M} \left( \frac{k_{\nu_i}+\kappa_{\nu_i}}{N} \right)} &  \text{if } l\geq l_{\tau_i} ,\\
% e^{j\theta_{i,\text{isi}}(l)} = e^{j2\pi \frac{l-M}{M} \left( \frac{k_{\nu_i}+\kappa_{\nu_i}}{N} \right)} & \text{otherwise}
% \end{cases} 
\end{align}
\end{theorem}
%\todo{I remove the notation $e^{j\theta_i(l)}$ in the expression. Can you please help me to remove them in the proofs too?}
%\todo{Do you think it can better if we first include the general result (similar (13) for the ideal case) for this case first (probably using expressions (56) and (64)). We then derive (24) (which is (19)'s counterpart), which incorporates only significant IDI terms only. By doing this, the results of both ideal and rectangular cases are more coherent and consistent. }
{\em Proof:}  The proof is relegated to the Appendix \ref{app_th_rect_dd}. \hfill $\blacksquare$ \\
{\color{black}Note that the approximation error in (\ref{conv_eq_rect}) is very small and it reduces by increasing $N$ (see (\ref{eq_new1}) in Appendix \ref{app_th_rect_dd}).}
Theorem \ref{th_rect_dd} implies that the ICI and ISI in time--frequency domain are converted to simple phase shifts in the delay--Doppler domain.
Moreover, from (\ref{conv_eq}) and (\ref{conv_eq_rect}), we can observe that the number of transmitted signals that affects a received signal is the  same for both ideal and rectangular pulse cases. The only difference is that the channel is shifted by an additional phase that depends on the location of the transmitted signal in the delay--Doppler plane (i.e., $k$ and $l$)
%\todo{It could be nice if we can include the equation for the case of no fractional Doppler.}

{\em Special channel model cases:}
Let us consider the above input-output expression (\ref{conv_eq_rect}) for the special cases mentioned in Sec. \ref{sec_special}. 

{\em i) Ideal channel}: The received signal becomes
\begin{align}
y[k,l] & = x[k,l],
\nonumber
\end{align}
which is the same as the ideal pulses case since the rectangular pulses satisfy the bi-orthogonal property in (\ref{bi-ortho}) when the channel is ideal (i.e., $\tau_{\rm max}=0$ and $\nu_{\rm max}=0$). This can be seen easily by observing (\ref{cross}) at $t=nT$ and $f=m\Delta f$. %\todo{\tiny{Elaborate more on why}} 

{\em ii) No fractional Doppler} (i.e., $\kappa_{\nu_i}=0,\forall i)$: Equation (\ref{conv_eq_rect}) simplifies to: 
%\todo{\tiny need too say the approx error is within $1/N$ }
\begin{align}
y[k,l] {\color{black}\approx} \sum_{i=1}^{P} h_i e^{j2\pi \left(\frac{l-l_{\tau_i}}{M}\right)  \frac{k_{\nu_i}}{N} }  \alpha_i(k,l)  x[[k - k_{\nu_i}]_N, [l - l_{\tau_i}]_M],
\nonumber
\end{align}
where
\begin{align*}
\alpha_i(k,l) & = 
\begin{cases}
1 &  l_{\tau_i} \leq l < M\\
\frac{N-1}{N} e^{-j 2 \pi \left(\frac{[k-k_{\nu_i}]_N}{N} \right)} &  0\leq l<l_{\tau_i}
\end{cases}
\end{align*}
In this case, IDI does not appear as in the case of ideal pulses.  
%\todo{\tiny{Should it be $l \geq l_{\tau_i}$, not $l \leq l_{\tau_i}$? Please check and compare with (25).} }
%\todo{please check if it includes the term $e^{-j2\pi\nu_i\tau_i}$ as in the ideal case. -- No it doesn't have that term}
%\todo{$k'$ should not be here. Please check. If that is the case, you need to re-compute $\alpha_i, \beta_i$ accordingly.}
% \begin{remark}
% \textcolor{blue}{{\it OFDM with zero CP over delay--Doppler channels}. Is it the case when $N=1$, the above analysis becomes (conventional) OFDM with zero CP over delay--Doppler channels? What needs to be change? 
% We can move the materials in Section V.B (is it OFDM with CP over delay--Doppler channels ?) (if they are included) and contrast OTFS with OFDM here. Section V is dedicated only to MP algorithm.  }
% \end{remark}
% \subsubsection{Input--output relationship}
% \subsubsection{Phase shifts}
\section{Message passing algorithm for joint interference cancellation and  detection}%\todo{\tiny REMINDER: Fix the probability normalization in MP. \\ 
% Dear Ravi, so far, we have maintained quite consistent use of notations. Can you please update (and improve) the notations for the MP section according to previous sections. }
We now propose a message passing (MP)  algorithm for OTFS using the input-output relation in (\ref{conv_eq}) (or (\ref{conv_eq_rect})). 
\subsection{Low-complexity MP detection algorithm}
The received signal in vectorized form can be written as
%\todo{\tiny Some references about the MP and the Gaussian approximation method, say messages are mean and variance}
\begin{align}
\vy = \mh \, \vx + \vz
\label{vec_form}
\end{align}
where $\vy = \{y[d]\}, \vz = \{z[d]\}\in \mathbb{C}^{NM \times 1}, \mh =\{{H}[d,c]\}\in \mathbb{C}^{NM \times NM},$ and $\vx=\{x[c]\} \in \mathbb{A}^{NM \times 1}$. The elements $\vy, \vx$, and $\mh$ are determined from (\ref{conv_eq}) (or (\ref{conv_eq_rect})) and $\vz$ is the noise vector. 
%The $(k+N l)$-th element of $\vy$ is ${y}[k+N l] = y[k,l],$ for $k =0 ,\cdots,N-1$, $l = 0,\cdots, M-1$.  
%The elements of $\vx$ and $\vz$ are similarly related to $x[k,l]$ and $z[k,l]$, respectively. 
Due to mod $N$ and mod $M$ operations in (\ref{conv_eq}),
we observe that only ${S} = \sum_{i=1}^P (2N_i+1)$ elements out of $NM$ are non-zero in each row and column of $\mh$. Recall that $P$ is the number of propagation paths. We can see that since $S$ is much smaller than $NM$, $\mh$ is a sparse matrix. Let $\mathcal{I}(d)$ and $\mathcal{J}(c)$ denote the sets of indexes with non-zero elements in the $d$-th row and $c$-th column, respectively, then $|\mathcal{I}(d)| = |\mathcal{J}(c)| = {S}$ for all rows and columns. 
%\todo{\tiny Should it be  $\mathcal{I}_{d}$ and $\mathcal{J}_c$ (see also figure) ? Why are they the same size? }  \
Note that although (\ref{vec_form}) applies to both ideal pulses case in (\ref{conv_eq}) and rectangular pulses case in (\ref{conv_eq_rect}), with different $\mh$'s, the number of non-zero elements $S$ in each row and column of $\mh$ remains  the same for both cases. This condition helps in compensating ICI and ISI of rectangular pulses with the same complexity detection algorithm of ideal pulses.  

Based on (\ref{vec_form}), we model the system as a sparsely-connected factor graph with $NM$ variable nodes corresponding to $\vx$ and $NM$ observation nodes corresponding to $\vy$. In this factor graph, each observation node $ {y}[d]$ is connected to the set of $S$ variable nodes $\{ {x}[c], c \in \mathcal{I}(d)\}$. Similarly, each variable node $ {x}[c]$ is connected to the set of $S$ observation nodes $\{ {y}[d], d \in \mathcal{J}(c)\}$. %\todo{\tiny A figure could be useful if enough space otherwise add a reference}

\begin{figure}
\begin{minipage}[b]{0.5\columnwidth}
\centering
\includegraphics[scale=0.75]{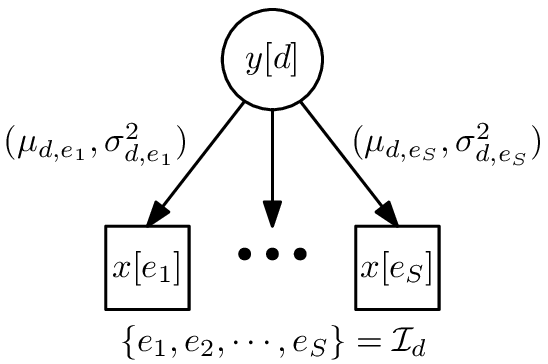}\\
{\footnotesize Observation node messages}
\end{minipage}%
\begin{minipage}[b]{0.5\columnwidth}
\centering
\includegraphics[scale=0.75]{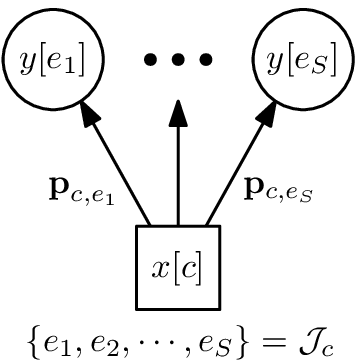}\\
{\footnotesize Variable node messages}
\end{minipage}
\caption{Messages in factor graph}
\label{mp_graph}
\end{figure}
%\todo{Please update Figure 2 using new notations}
\iffalse
\begin{figure}
\hspace{-5mm}
\subfloat[Observation node messages] {\includegraphics[scale=0.5]{obs_var2}}
\hspace{5mm}
\subfloat[Variable node messages] {\includegraphics[scale=0.5]{var_obs2}}
\caption{\normalsize Messages in factor graph}
\label{mp_graph}
\end{figure}
\fi

From (\ref{vec_form}), the joint maximum a posterior probability (MAP) detection rule for estimating the transmitted signals is given by
\begin{equation}
\widehat{{\bf x}} = \argmax_{ {{\bf x}} \in \mathbb{A}^{NM \times 1}} \, \Pr \left( \vx \;\!\big|\;\!  \vy,\mh \right),
\nonumber
\end{equation}
which has a complexity exponential in $NM$. Since the joint MAP detection can be intractable for practical values of $N$ and  $M$, we consider the symbol-by-symbol MAP detection rule for $c=1,\hdots,NM$ %\todo{since $c$ is the vector index, so $ c=1,\hdots,NM$? in earlier $x[k,l]$, we use from 0 to $N-1$, 0 to $M-1$}
\begin{align}
\widehat{x}[c] & = \argmax_{ a_j \in \mathbb{A}} \, \Pr \left( x[c] = a_j \;\!\big|\;\! \vy,\mh \right)
 \\
& = \argmax_{ a_j \in \mathbb{A}} \, \frac{1}{Q} \Pr \left( \vy \;\!\big|\;\! x[c] = a_j,\mh \right) \label{sym_map1} \\
& \approx \argmax_{ a_j \in \mathbb{A}} \prod_{d \in \mathcal{J}_c} \Pr \left( y[d] \;\!\big|\;\!  x[c] = a_j,\mh \right).
\label{sym_map}
\end{align}
%\todo{\tiny there is another simplification when you do the first step you remove Pr$(x_i)$ by assuming they are all equal to $1/Q$ but this is true only in the first iteration}
%\todo{Explain little more why from (24) to (25).}
In  (\ref{sym_map1}), we assume all the transmitted symbols $a_j \in \mathbb{A}$ are equally likely and in (\ref{sym_map}) we assume the components of $\vy$ are approximately independent for a given $x[c]$, due to the sparsity of $\mh$. That is, we assume the interference terms $\zeta_{d,c}^{(i)}$ defined in (\ref{gau_app}) are independent for a given $c$. 
%\todo{Can you elaborate this statement? }. %\todo{\tiny Is it possible to characterize the cycle lengths in the factor graph? do cycles introduce error floors?}
In order to solve the approximate symbol-by-symbol MAP detection in (\ref{sym_map}), we propose a MP detector which has a linear complexity  in $NM$. 
Similarly to \cite{mp_ga}, for each $y[d]$, a variable $x[c]$ is isolated from the other interference terms, which are then approximated as Gaussian noise with an easily computable mean and variance. 

In the MP algorithm, the mean and variance of the interference terms are used as messages from observation nodes to variable nodes. On the other hand, 
the message passed from a variable node ${x}[c]$ to the observation nodes ${y}[d], d \in \mathcal{J}(c)$, is the probability mass function (pmf) of the alphabet
${\bf p}_{c, d} = \left\{ p_{c, d}(a_j) |  a_j \in \mathbb{A} \right\}$. %\todo{Better ${\bf p}_{c, d}$ not ${\bf p}_{c d}$. Please change other places too.}
Fig. \ref{mp_graph} shows the connections and the messages passed between the observation and variable nodes.
The MP algorithm is described in {\bf{Algorithm 1}}. \\%\todo{\tiny $ite$ is ugly}
%\todo{Reconsider other presentations for the MP algorithm. I think the current presentation is not so clear. }
  \begin{algorithm}
  Input: Received signal $\vy$, channel matrix  $\mh$.\\ 
  Initialization: pmf ${\bf p}_{c, d}^{(0)} = 1/Q, c =1,\cdots,NM, d \in \mathcal{J}(c)$, iteration count $i=1$. 

    \Repeat{Stopping criteria }{
    - Observation nodes $y[d]$  compute the means $\mu_{d,c}^{(i)}$ and variances $(\sigma^{(i)}_{d,c})^2 $ of Gaussian random variables $\zeta_{d,c}^{(i)}$ using  ${\bf p}^{(i-1)}_{c,d}$ and pass them to variables nodes $x[c], c \in \mathcal{I}(d)$. \\ 
    - Variable nodes $x[c]$ update  ${\bf p}^{(i)}_{c,d}$ using $\mu_{d,c}^{(i)}$, $(\sigma^{(i)}_{d,c})^2 $, and  ${\bf p}^{(i-1)}_{c,d}$ and  pass them to observation nodes $y[d], d \in \mathcal{J}(c)$. \\    
     - Compute convergence indicator $\eta^{(i)}$. \\
     - Update the decision on the transmitted symbols $\widehat{x}[c], c=1,\hdots,NM$ if needed. \\ 
     - $i=i+1$
    } 
    Output: The decision on transmitted symbols $\widehat{x}[c].$
  \caption{MP algorithm for OTFS symbol detection}
\end{algorithm}
The details of the steps in iteration $i$ in the MP algorithm are detailed below.  
%{\bf Step 1}: Initialize iteration index $i=1$ and ${\bf p}_{c, d}^{(0)} = 1/Q$ for 
% $c = 0,\cdots,NM-1$, and $d \in \mathcal{J}_c$.   
 
%{\bf Step 2}: Messages are passed from the observation nodes to the variable nodes.
%The message passed from $ {y}[d]$ to $ {x}[c]$ is the pdf of a Gaussian random variable $\zeta_{d,c}^{(i)}$ %with mean $\mu_{dc}^{(i)}$ and variance %$(\sigma_{dc}^{(i)})^2$, 
%which can be computed as: %the received value $ {y}[d]$ by writing it as%\todo{\tiny Define the noise+interference term $\zeta$ and simplify the rest}
{\bf Message passings from observation nodes $y[d]$ to variable nodes $x[c], c \in \mathcal{I}(d)$}:
The mean $\mu_{d,c}^{(i)}$ and variance $(\sigma^{(i)}_{d,c})^2 $ of the interference, approximately modeled as  a Gaussian random variable $\zeta_{d,c}^{(i)}$ defined as: 
\begin{align}
 {y}[d] =  {x}[c]  {H}[d,c] + \underbrace{ \sum_{e \in \mathcal{I}(d), e \neq c}  {x}[e]  {H}[d,e] +  {z}[d]}_{\zeta_{d,c}^{(i)}},
 \label{gau_app}
\end{align}
can be computed as:  
\begin{align}
\mu_{d,c}^{(i)} & =  \sum_{e \in \mathcal{I}(d), e \neq c} \sum_{j=1}^{Q} p_{e,d}^{(i-1)}(a_j) a_j  {H}[d,e], 
\label{mean_com}
\end{align}
and 
\begin{eqnarray}
 (\sigma^{(i)}_{d,c})^2  = 
  \sum_{e \in \mathcal{I}(d), e \neq c} \! \left( \sum_{j=1}^{Q} p_{e,d}^{(i-1)}(a_j) |a_j|^2 | {H}[d,e]|^2   
- \left|\sum_{j=1}^{Q} p_{e,d}^{(i-1)}(a_j) a_j  {H}[d,e]\right|^2  \right) \! +  \sigma^2 .
 \label{var_com}
\end{eqnarray}
%Further, we assume that transmitted symbols are i.i.d. and independent from the noise.\\
{\bf Message passings from variable nodes $x[c]$ to observation nodes $y[d], d \in \mathcal{J}(c)$}:  The pmf vector ${\bf p}^{(i)}_{c,d}$ can be updated as: 
\begin{align}
p_{c,d}^{(i)}(a_j) & = \Delta \cdot  {\tilde{p}}_{c,d}^{(i)} (a_j) + (1-\Delta) \cdot p_{c,d}^{(i-1)} (a_j), a_j \in \mathbb{A}
\label{prob_com}
\end{align}
where $\Delta \in (0,1]$ is the {\em damping factor} used to improve the performance by controlling the convergence speed \cite{damp}, and %\todo{\tiny what performance is improved? Does it trade off performance and convergence speed?}  
\begin{align}
{\tilde{p}}_{c,d}^{(i)}(a_j) & \propto \prod_{e \in \mathcal{J}(c), e \neq d} \Pr \left( y[e] \Big| x[c] = a_j,\mh \right) \nonumber \\
& = \prod_{e \in \mathcal{J}(c), e \neq d} \frac{\xi^{(i)}(e,c,j)}{\sum_{k=1}^{Q} \xi^{(i)}(e,c,k)}~, 
\label{pcdaj}
\end{align}
where $\xi^{(i)}(e,c,k) = \exp \left( \frac{-\left| {y}[e] - \mu_{e,c}^{(i)} -  {H}_{e,c}a_k  \right|^2} {(\sigma^{(i)}_{e,c})^2} \right)$.
%\todo{\tiny shall we keep the details of the normalization for the journal?} 
%Note that (\ref{pcdaj}) excludes the information of ${y}[d]$.
% \todo{\tiny{I'm not sure whether we should refer it as `convergence rate' since I believe that convergence rate in iterative algorithm means something else. I would think 'convergence indicator' may be better.}}

 {\bf{Convergence indicator}}: Compute the convergence indicator $\eta^{(i)}$ as
 \begin{align}
 \eta^{(i)} & = \frac{1}{NM}\displaystyle\sum_{c=1}^{NM} {\mathbb I}\left(\max_{a_j \in \mathbb{A}} \, \, p_c^{(i)}(a_j) \geq 1-\gamma \right),
 \label{conv_ind}
 \end{align}
 for some small $\gamma >0$ and where 
 \begin{align}
p_c^{(i)}(a_j) = \prod_{e \in \mathcal{J}(c)} \frac{\xi^{(i)}(e,c,j)}{\sum_{k=1}^{Q} \xi^{(i)}(e,c,k)}
\label{pcaj}
\end{align}
 and ${\mathbb I}(\cdot)$ is an indicator function. 
%\todo{Should it be $\ge 1-\gamma$?}
%\todo{Should not explicitly mention .99. Just saying some generic umber close to 1, which controls the convergence of the MP algorithm etc.} 

{\bf{Update decision}}: If $\eta^{(i)} > \eta^{(i-1)}$, then we update the decision of the transmitted symbol as 
\begin{equation}
\widehat{x}[c]  = \argmax_{a_j \in \mathbb{A}} \, \, p_c^{(i)}(a_j), ~~ c=1,\cdots,NM.
\label{upd_dec}
\end{equation}
We update the decision on the transmitted symbols only when the current iteration can provide better estimates than the previous iteration. 
 %\todo{It seems $\widehat{x}^{(i)}[c]$ is not used in Steps 2, and 3. If so, then we don't need to determine $\widehat{x}^{(i)}[c]$ for each iteration, probably needed in Step 5 only.}
 
{\bf{Stopping criteria.}} The MP algorithm stops when at least one of the following conditions is satisfied.
\begin{enumerate}
\item $\eta^{(i)} = 1$
\item $\eta^{(i)} < \eta^{(i^*)}-\epsilon$, where $i^*$ is the iteration index from $\{1,\cdots,(i-1)\}$ for which $\eta^{(i^*)}$ is maximum
\item Maximum number $n_{iter}$ of iterations is reached.
\end{enumerate}
We select $\epsilon=0.2$ to disregard small fluctuations of $\eta$. 
Here, the first condition occurs in the best case, where all the symbols have converged. The second condition is useful to stop the algorithm if the current iteration provides a  worse decision than the one in previous iterations. 

%{\bf Output}: The final decision about the transmitted symbols is $\widehat{x}[c]$ for $c \in \{0,\cdots,NM-1\}$.   
%  \begin{equation*}
% \widehat{x}[c]  =  \widehat{x}^{(i^*)}[c],  ~~ c \in \{0,\cdots,NM-1\}.
% \end{equation*}
% \begin{equation*}
% \widehat{x}[c]  = \argmax_{a_j \in \mathbb{A}} \, \, p_c(a_j), ~~ c \in \{0,\cdots,NM-1\}.
% \end{equation*}
% where 
% \[p_c(a_j) = \prod_{e \in \mathcal{J}_c} \Pr \left( y[e] \Big| x[c] = a_j,\mh \right).
% \]

\begin{remark}
{\em Complexity of the proposed MP algorithm.}
%The complexity of the algorithm is the product of the number of iterations and per iteration complexity. 
The complexity of one iteration involves the computation of (\ref{mean_com}), (\ref{var_com}), (\ref{prob_com}), (\ref{conv_ind}), and (\ref{upd_dec}).
% and (\ref{prob_com}), 
%where each computation has a complexity of the order $\mathcal{O} (NM S Q)$. 
More specifically, each of (\ref{mean_com}), (\ref{var_com}), and (\ref{prob_com}) \footnote{In computing (\ref{pcdaj}), first we find the $p_c^{(i)}(a_j)$ in (\ref{pcaj}) which requires $\mathcal{O} (NM Q)$ complexity and then we obtain (\ref{pcdaj}) by dividing (\ref{pcaj}) with the term related to $e=d$ for all $d$, which requires $\mathcal{O} (S)$ complexity for each $c$. Hence, the over all complexity of (\ref{pcdaj}) becomes $\mathcal{O} (NM S Q)$.}, has a complexity order $\mathcal{O} (NM S Q)$. Furthermore, (\ref{conv_ind}) and (\ref{upd_dec}) can be computed with a complexity order $\mathcal{O} (NM Q)$ \footnote{The computation of (\ref{conv_ind}) and (\ref{upd_dec}) require to find the maximum element out of $Q$ elements for each $c$. As (\ref{pcaj}) is already computed for (\ref{pcdaj}), finding the maximum element requires $\mathcal{O} (Q)$ complexity for each $c$, which leads to an overall complexity of $\mathcal{O} (NM Q)$ to compute (\ref{conv_ind}) and (\ref{upd_dec}).}.  
Therefore, the overall complexity order is $\mathcal{O} (n_{iter}N M S Q)$. In simulations, we observed that the algorithm converges typically within 20 iterations (i.e., see  Figure \ref{delta_fig} in the illustrative result section for more references). We conclude that the IDI analysis, which includes the smart approximation of IDI, to exploit the sparsity of the delay-Doppler channel representation is a key factor in reducing the complexity of the detector (due to relatively small $S$). 
%\todo{This is not really precise. Probably due to our IDI analysis and 'smart' approximation to exploit the channel sparsity. Consider rephrasing.}
The memory requirement is dominated by the storage of $2NMS Q$ real values for ${\bf p}_{c,d}^{(i)}$ and ${\bf p}_{c,d}^{(i-1)}$. In addition, we have the messages $(\mu_{d,c}^{(i)},(\sigma_{d,c}^{(i)})^2)$, requiring $NMS$ complex values and $NMS$ real values, respectively.
\end{remark}
\subsection{Application of MP detection algorithm for OFDM over delay--Doppler channels}
Later, we will compare the performance of OTFS and OFDM over delay--Doppler channels. In this section, we demonstrate that 
it is also possible to apply the above MP algorithm to OFDM to compensate the Doppler effects. 

% \textcolor{blue}{The OFDM system can be illustrated by the inner dashed box in Fig.~\ref{sys_fig}, i.e., the Time-Frequency domain.
% Specifically, the Heisenberg Transform module is replaced by IFFT, cyclic prefix (CP) addition, serial-to-parallel and digital-to-analog conversion, and the Wigner Transform module is substituted with analog-to-digital, parallel-to-serial, CP removal and FFT operation. Also, as mentioned in {\it Remark 1}, for OFDM systems, $N$ is set to 1.} %In the following, we omit the details due to space limitation.

Consider OFDM system with OFDM symbol of duration $T$ and $M$ subcarriers. Hence, the received signal  and noise  are sampled at $T/M$. Then, the frequency-domain signal after fast Fourier transform (FFT) operation is given by:
\begin{align}
\mathbf{{y}} = \mathbf{W}\mathbf{{H}}_t\mathbf{W}^H\mathbf{{x}} + \mathbf{{z}}
\label{vec_form_ofdm}
\end{align}
where $(\cdot)^H$ denotes Hermitian transpose, $\mathbf{W}$ is $M$-point FFT matrix, and $\mathbf{{x}} \in \mathbb{A}^{M\times 1}$ is the transmitted OFDM symbol.  
% and
% \begin{align*}
% \mathbf{{y}}=\mathbf{W}\mathbf{{r}}%~~~\mbox{and}~~~\mathbf{{v}}=\mathbf{W}\overline{\mathbf{v}}
% \end{align*}
% where $\mathbf{{r}} \in \mathbb{C}^{M\times 1}$, ${\mathbf{v}} \in \mathbb{C}^{M\times 1}$, 
% and the $m^{th}$ element of $\mathbf{r}$ and ${\mathbf{v}}$ are attained by sampling the received signal $r(t)$ and noise $v(t)$ in (\ref{delay-doppler}) at $T/M$, respectively.
The elements of time-domain channel matrix $\mathbf{{H}}_t = \{{H}_t[p,q]\}$ are given in \cite{Zhao} as
\[
{H}_t[p,q]\!=\!\sum_{i=1}^{P}h_i\delta\!\left[\!\left[p-q-\frac{\tau_iM}{T}\right]_{\!M}\!\right]\!e^{j\frac{2\pi (q-1)\nu_i}{M}}
\]
for $p, q =1,\hdots,M$. 
%\todo{Should it be the summation over all paths? }
Using the ${M\times M}$ frequency-domain channel matrix $\mathbf{{H}}_{\text{ofdm}} = \mathbf{W}\mathbf{{H}}_t\mathbf{W}^H$, we can re-write (\ref{vec_form_ofdm}) as: 
\begin{align}
\mathbf{{y}} = \mathbf{{H}}_{\text{ofdm}}\mathbf{{x}} + \mathbf{{z}}. \label{vec_form_ofdm1}
\end{align}
Since (\ref{vec_form_ofdm1}) has the form similar to (\ref{vec_form}), the MP previously developed for OTFS can also be applied for OFDM symbol detection.
% \begin{align*}
% \mathbf{\overline{H}_t}\triangleq~~~~~~~~~~~~~~~~~~~~~~~~~~~~~~~~~~~~~~~~~~~~~~~~~~~~~~~~~~~~~~~~\\
% \begin{bmatrix}
% b_{1,0}&&&&b_{L,M-1}~\dots~b_{2,M-1}\\
% b_{2,0}&b_{1,1}&&&\ddots~~~~~~\vdots \\      
% \vdots &\ddots & &&~~~~~~~~~~~b_{L-1,M-L-1}\\ 
% b_{L,0}&\dots&b_{1,L-1} \\
% &\ddots &\ddots&\ddots \\
% &&\ddots &\ddots &\ddots~~~~~~~~~~~~~~\\
% &&&b_{L,M-L}&~~\dots~~~~~~b_{1,M-1}\\
% \end{bmatrix},
% \end{align*}
% where $L=\tau_{max}$ and $b_{p,q}=a_i\delta(p-\frac{\tau_iM}{T})e^{j\frac{2\pi q\nu_i}{M}}$ from (\ref{del_dop}).
We note that  $\mathbf{{H}}_{\text{ofdm}}$ is {\em diagonally dominant} and the values of off-diagonal elements in each row decay as we move away from the diagonal entry as explained in \cite{Zhao}. %\todo{\tiny{we should explain or elaborate little more on the reasons or provide reference because it is not so clear to see.}}. 
Hence, the $\mathbf{{H}}_{\text{ofdm}}$ matrix is also sparse enabling the use of the proposed low-complexity MP detection algorithm. 

% {\remark} From (\ref{conv_eq}) and (\ref{vec_form_ofdm}), we can observe the effects of channel gain on the transmitted symbols are different in OTFS and OFDM. In OTFS, all the transmitted symbols experience the same channel gain (independent of $k$ and $l$), whereas in OFDM, the channel gains are distinct at different subcarriers because of the FFT operation on $\mh_t$. %\todo{\tiny{Not so clear to see. I think it is easier to see from (13), not (19) -- In (13) gain still depends on $k,l$ only (19) explains the independence, please see now, I slightly changed the (19)}}

%\todo{\tiny Comment it is not a sparse graph as OTFS}
%and the performance comparisons are shown in the next section.

\section{Illustrative results and discussions}

% \begin{enumerate}
% \item 
% Fractional effect on OTFS no correction vs correction of 1, 5, 10 subcarriers (IDI)
% %\item
% %Figure to show effect of damping factor in MP
% \item
% OTFS vs OFDM 4-QAM  at  30km/h, 120km/h, 500km/h   Message is that Doppler completely removed
% \item
% OTFS vs OFDM 16   at   120km/h, 500km/h   Message is that Doppler completely removed
% \end{enumerate}

In this section, we simulate the uncoded bit-error-rate (BER) performance of OTFS and OFDM over delay-Doppler channels. In particular, first we study the BER performance of OTFS for ideal pulses with the number of interference terms due to IDI $N_i$ and MP parameter $\Delta$. Next, we study the BER performance of OTFS with ideal pulses and rectangular pulses, and its comparison with OFDM. 

All relevant simulation parameters are given in Table \ref{tab_sp}. 
For both OTFS and OFDM systems, Extended Vehicular A model\cite{LTE} is adopted as the channel model for the path delays, and the Doppler shift of the $i-$th path is generated using
%Jakes' model \cite{Jakes} is adopted for the channel Doppler model \todo{\tiny we discussed this before, this is not Jakes model, suggest to remove the Jake's model Claim}, where the Doppler shift of the $i$-th path is given as
\[
\nu_i = \nu_{\mbox{max}}\cos(\theta_i),
\]
where $\theta_i\sim \mathcal{U}(0,\pi)$ is uniformly distributed.
%Note that OFDM system has negligible 4\% spectral efficiency %loss due to CP. 
In order to obtain BER values, we consider $3\times 10^{4}$ different channel realizations in the Monte-Carlo simulations.

\begin{table}[t]
\centering
  \begin{tabular}{ | r | p{2cm} | }
    \hline
    Parameter & Value  \\ \hline
    Carrier frequency  & 4 GHz  \\ \hline
    No. of subcarriers ($M$) & 512  \\ \hline
    No. of OTFS symbols ($N$) & 128 \\ \hline
    Subcarrier spacing & 15 KHz \\ \hline
    Cyclic prefix of OFDM & 2.6 $\mu$s\\ \hline
    %No. of OFDM symbols & 120 \\ \hline
    Modulation alphabet & 4-QAM \\ \hline
    %Modulation alphabet for OFDM & 4-QAM for 112 symbols and 16-QAM for 8 symbols \\ \hline
    UE speed (Kmph) & 30, 120, 500\\ \hline
    %Channel Doppler model & Jakes' model\cite{Jakes} \\ \hline
    %Channel delay model & Extended Vehicular A model\cite{LTE} \\ \hline
    Channel estimation & Ideal \\ \hline
  \end{tabular}
  \vspace{2mm}
\caption{Simulation Parameters}
\vspace{-2mm}
\label{tab_sp}
\vspace{-5mm}
\end{table}
%\todo{Since Monte Carlo simulations are used to obtain BER. How many channel realizations did you use? Please specify.} 

\begin{figure}
\centering
\includegraphics[width=3in,height=2.2in]{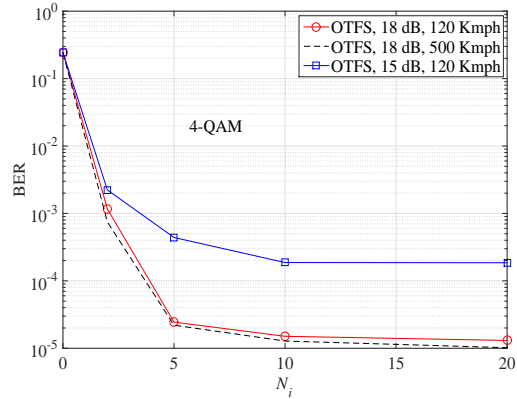}%clip=true,trim=1.5mm 1mm 1.4cm 7mm
\caption{The BER performance of OTFS for different number of interference terms $N_i$ with $4$-QAM.}
\label{sim1}
\end{figure}

We first demonstrate the effects of IDI in OTFS. Fig. \ref{sim1} shows the BER performance of OTFS system with ideal pulses using the proposed MP detector for different number of IDI interference terms $N_i$ with $4$-QAM signaling over the delay--Doppler channel with different Doppler frequencies (UE speeds of $120, 500$ Kmph) and SNRs. Note that ICI and ISI are not present for the ideal pulses case. We consider the same $N_i$ for all paths. We can see that there is a significant BER improvement when $N_i$ increases from $0$ to $10$ and saturation thereafter. Note that $N_i=0$ corresponds to the case when IDI is not taken into account. The results imply that fewer neighboring interference terms are sufficient to consider in the MP algorithm (e.g. $N_i=10$) without incurring  performance loss. We also observe that if IDI is not taken into account at all or an insufficient number of IDI terms is considered (i.e., $N_i \le 5$), the BER performance worsens significantly. These observations demonstrate the importance of our previous IDI analysis. Also, note that for $\text{SNR}=18\!$ dB, the BER performances of OTFS at different Doppler frequencies are similar. Later, we will demonstrate that our proposed MP algorithm can effectively compensate for a wide range of channel Doppler variations. 
%\todo{It could be better if we include 1 more plot for a different SNR value or velocity.}

\begin{figure}
\centering
\includegraphics[width=1.7in,height=1.8in,clip=true,trim=1.5mm 2mm 1.5cm 6mm]{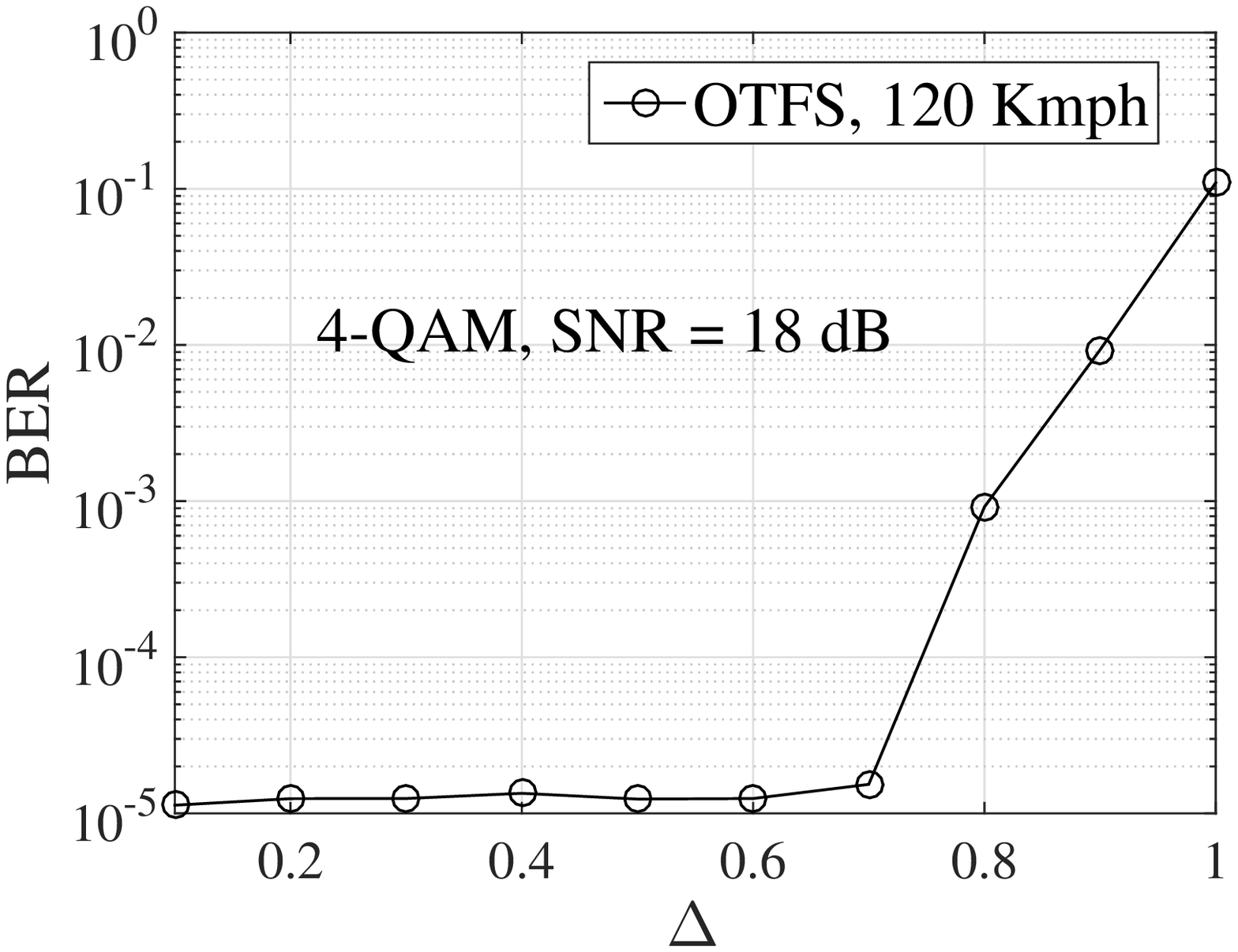}
\includegraphics[width=1.7in,height=1.8in,clip=true,trim=5mm 2mm 1.5cm 6mm]{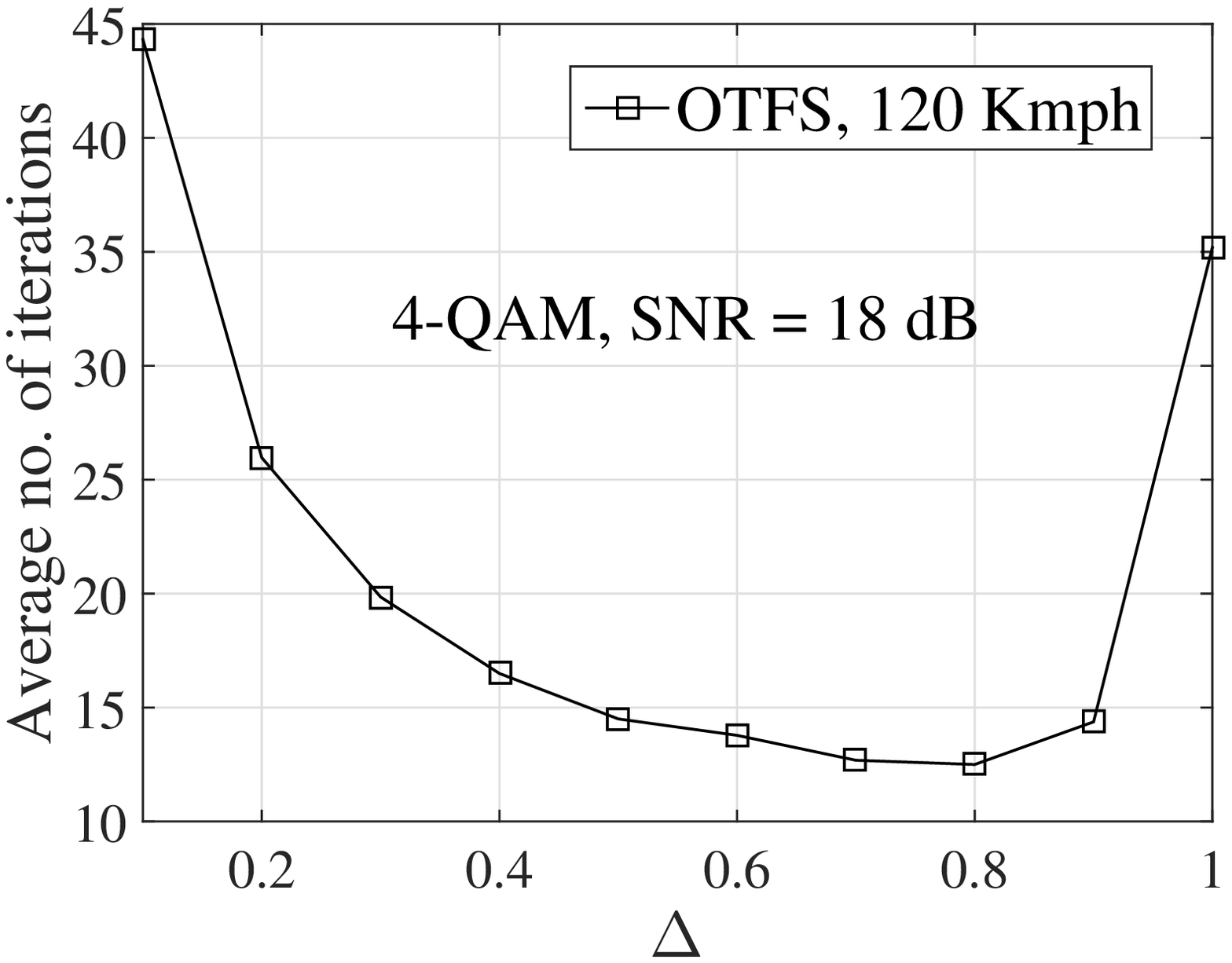}
\caption{The variation of BER and average no. of iterations with $\Delta$.}
\label{delta_fig}
\end{figure}

In Fig. \ref{delta_fig}, we illustrate the BER and average number of iterations of OTFS system with ideal pulses using the MP algorithm, when UE speed is $120$ Kmph. We vary the damping factor $\Delta$ for $N_i = 10$. We consider $4$-QAM signaling and SNR $= 18$ dB. We observe that, when $\Delta \leq 0.7$, the BER  remains almost the same, but deteriorates thereafter. 
Further, when $\Delta = 0.7$, the MP algorithm converges with the least number of iterations. Hence, we choose $\Delta=0.7$ as the optimum damping factor in terms of performance and complexity.
%\todo{Again, it could be better if we include 1 more plot for a different SNR value or velocity.}

\begin{figure}
\centering
\includegraphics[width=3in,height=2.2in]{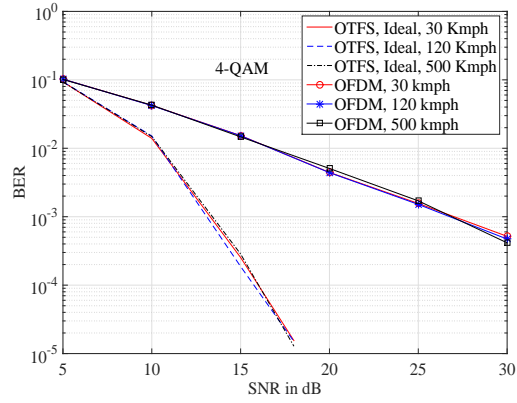}%,clip=true,trim=1.5mm 1mm 1.5cm 8mm
\caption{The BER performance comparison between OTFS with ideal pulses and OFDM systems at different Doppler frequencies.}
\label{sim2}
\end{figure}

In Fig. \ref{sim2}, we compare the BER performance of OTFS with ideal pulses and OFDM systems using $4$-QAM signaling over the delay-Doppler channels of different Doppler frequencies (UE speeds of $30,120,500$ Kmph). We observe that OTFS outperforms OFDM by approximately $15$ dB at BER of $10^{-4}$ thanks to the constant channel gain over all transmitted symbols in OTFS, whereas in OFDM, the overall error performance is limited by the subcarrier(s) experiencing the worse channel conditions.  Moreover, OTFS exhibits the same performance at different Doppler frequencies thanks to the IDI cancellation provided by the MP detector. Similar behavior applies to OFDM, since the ICI can be removed by the MP detector. We can conclude that the performance of OTFS under the proposed MP algorithm is  robust to Doppler variations and is much better than that of OFDM. 

\begin{figure}
\centering
\includegraphics[width=3.7in,height=2.7in,clip=true,trim=1.5cm 0mm 0mm 0mm]{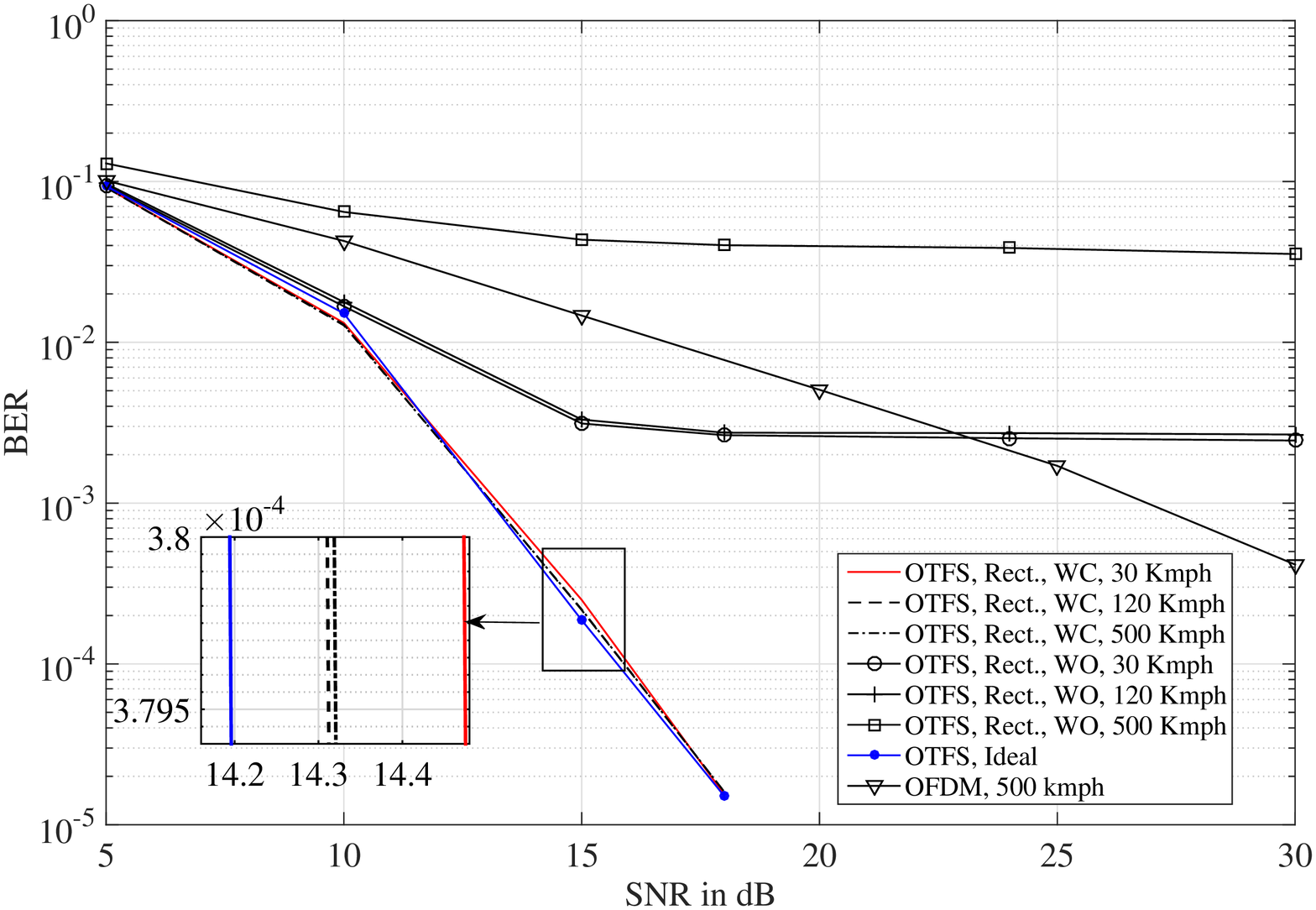}
\caption{The BER performance of OTFS with rectangular and ideal pulses at different Doppler frequencies for $4$-QAM.}
\label{sim3}
\end{figure}

\begin{figure}
\centering
\includegraphics[width=3in,height=2.2in]{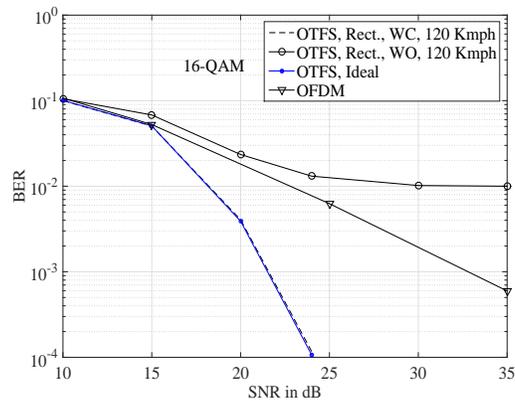}%,clip=true,trim=1.5mm 1mm 1.5cm 8mm
\caption{The BER performance of OTFS with rectangular and ideal pulses for $16$-QAM.}
\label{sim4}
\end{figure}

Fig. \ref{sim3} shows the BER of OTFS with rectangular pulses using $4$-QAM signaling for two scenarios: one with ICI and ISI cancellations (WC) and the other without (WO).  
% using the two different channels in the MP detection:
% \begin{enumerate}
% \item ideal pulses channel in (\ref{conv_eq}), i.e., without ICI and ISI cancellation (WO)
% \item rectangular pulses channel in (\ref{conv_eq_rect}), i.e, with ICI and ISI cancellation (WC)
% \end{enumerate}
In the second scenario, we observe that OTFS with rectangular pulses present an error floor incurred by the ICI and ISI. The performance degradation becomes more severe at high Doppler (e.g., $500$ Kmph) due to large ICI and ISI. On the other hand, OTFS with rectangular pulses approaches the  BER performance of OTFS with ideal pulses, when ISI and ICI are  mitigated. Moreover, we can see that the proposed MP algorithm can effectively remove ISI and ICI and thus OTFS performance remains almost constant regardless of the Doppler frequencies. These results show that it is possible to achieve the performance of OTFS with ideal waveforms under any Doppler frequencies  even with the more practical rectangular waveforms by using our  MP algorithm together with appropriate IDI, ICI and ISI cancellation. Last, we can see that while OTFS performance is not affected by (high) Doppler values, OFDM performance incurs error floor on channels with high Doppler frequencies. 

In Fig. \ref{sim4}, we compare the BER performance of OTFS and OFDM at a Doppler of $120$ Kmph using $16$-QAM signaling. We observe that OTFS with ICI and ISI cancellation outperforms OFDM by $11$ dB at BER = $10^{-3}$. We also simulate OTFS at different Doppler frequencies of $30$ and $500$ Kmph and we observe the BER performances are similar to that of $120$ Kmph.

\section{Conclusion}
In this paper, we have analyzed the input--output relation describing OTFS mod/demod over delay--Doppler channels. We have studied in detail the cases of ideal waveforms and  rectangular waveforms. In particular, we have characterized the  inter-Doppler interference (IDI), inter-carrier interference (ICI), and  inter-symbol interference (ISI) using sparse representation of the channel in the delay--Doppler domain. A low-complexity yet efficient message passing (MP) algorithm for joint IC and symbol detection was proposed, which is suitable for large-scale OTFS with inherent channel sparsity. In the MP algorithm, the ISI and ICI can be canceled by using appropriate phase shifting, while the IDI can be mitigated by accounting for a small number of significant interference terms only. The proposed MP algorithm can effectively compensate for a wide range of channel Doppler spreads. Moreover, we have demonstrated that it is possible to achieve the performance of OTFS with ideal yet {\it non-realizable} waveforms  using {\it practical} rectangular waveforms. Through simulations, we have shown that  OTFS has significant error performance gains over OFDM under various channel conditions.

\section*{Acknowledgement}
This research work is support by the Australian Research Council under Discovery Project ARC DP160100528. Simulations were undertaken with the assistance of resources and services from the National Computational Infrastructure (NCI), which is supported by the Australian Government. 

%\useRomanappendicesfalse
\appendices

% \begin{figure}
% \centering by $h_1$ and $h_2$ and applied in cascade to a waveform $g(t)$, we have: 
% \beqn
% \Pi_{h_2}(\Pi_{h_1}(g(t))) = \Pi_{h}(g(t)),
% \eeqn 
% where $h(\tau,\nu) = h_2(\tau,\nu) \ast_{\sigma}h_1(\tau,\nu)$ is the twisted convolution of $h_1(\tau,\nu)$ and $h_2(\tau,\nu)$ defined as: 
% \begin{align}
% & h_2(\tau,\nu) \ast_{\sigma}h_1(\tau,\nu) \nonumber \\  
% & = \int \int h_2(\tau',\nu') h_1(\tau - \tau',\nu -  \nu') e^{j2\pi\nu' (\tau -\tau')} d\tau' d\nu' \label{pro1_1} \\
% & = \int \int h_1(\tau',\nu') h_2(\tau - \tau',\nu -  \nu') e^{j2\pi(\nu-\nu') \tau} d\tau' d\nu'.
% \label{pro1_2}
% \end{align} 
% }
% {\em Proof:}  The proof is given in equations (\ref{pr1_1}) to (\ref{pr1_5}).  Starting from the is the direct expansion of Heisenberg operators (\ref{pr1_1}), we obtain (\ref{pr1_2}) by change of variables $\tau+\tau'=x$ and $\nu+\nu'=y$ in the inner integration. By exchanging order of integration in  (\ref{pr1_3}), we obtain the final expression (\ref{pr1_5}) of the Heisenberg transform  parameterized by twisted convolution of $h_2$ and $h_1$.  
% Finally, note that (\ref{pro1_2}) is obtained from  (\ref{pro1_1}) by the change of variables $\tau-\tau'\rightarrow \tau'$ and $\nu-\nu' \rightarrow \nu'$.

\section{Proof of Theorem \ref{thr_freq_time}: OTFS Input--Output Relation in Time--Frequency Domain }
\label{app_thr_freq_time}

%Let us consider the received signal after Wigner transform $Y(t,f)$,
% \begin{align*}
% Y(t,f) & = h(t,f) \ast_{\sigma} X[n,m] \ast_{\sigma} A_{g_{rx},g_{tx}}(t,f), \nonumber \\
% & = f(t,f) \ast_{\sigma} A_{g_{rx},g_{tx}}(t,f),
% \end{align*}
% where $f(t,f) = h(t,f) \ast_{\sigma} X[n,m]$. The value of $Y(t,f)$ can be expanded by using $f(t,f)$ as in (\ref{eq_f}) and (\ref{eq_Y}) respectively.
% %\todo{Should it be $h(\tau,\nu)$? Also, I think the first equation in (53) needs proof? -- to maintain (t,f) in Y, here we need (t,f) and I will write this proof as a lemma later.}
% The sampled version of $Y(t,f)$, i.e., $Y[n,m]$ in (\ref{eq_disY}) is derived from $Y(t,f)$ in (\ref{eq_Y}) by first change of variables $t'-n'T \rightarrow \tau$ and $f'-m'\Delta f \rightarrow \nu$ and then substituting $t=nT$ and $f=m\Delta f$. Finally, the result in (\ref{eq_freq_time}) is obtained by considering the value in square brackets in (\ref{eq_disY}) as $H_{n,m}[n',m']$, which completes the proof. \hfill$ \blacksquare$

The received signal after Wigner transform $Y(t,f)$, from (\ref{delay-doppler}), can be written as in (\ref{eq_th1}). It can be further expanded as in (\ref{eq_th2}) and (\ref{eq_th3}) using the transmitted signal $s(t)$ in (\ref{otfs1}) and some re-ordering of summations and integrations. Therefore, the sampled version of $Y(t,f)$, i.e., $Y[n,m]$, can be written as
\begin{align*}
Y[n,m] & = \sum_{n'=0}^{N-1}\sum_{m' = 0}^{M-1}   X[n',m'] H_{n,m}[n',m'],
\end{align*}
where $H_{n,m}[n',m']$ is given in (\ref{eq_th4}). By applying the change of variable $t'-\tau-n'T \rightarrow t''$ in the inner integral and some simple algebraic calculations, we can write $H_{n,m}[n',m']$ as in (\ref{eq_th5}) and (\ref{eq_th6}), respectively. Finally, we obtain $H_{n,m}[n',m']$ as in (\ref{eq_th7}), by replacing the square bracket in (\ref{eq_th6}) with cross-ambiguity function in (\ref{cross}), which completes the proof.

%\begin{figure*}
% ensure that we have normalsize text
\normalsize
% Store the current equation number.
%\setcounter{MYtempeqncnt}{\value{equation}}
%\setcounter{equation}{4}
\begin{align}
& Y(t,f)  = \int_{t'} g_{\text{rx}}^*(t'-t) \left[\int_{\tau} \int_{\nu} h(\tau,\nu) s(t'-\tau) e^{j2\pi\nu(t'-\tau)} d\tau d\nu \right] e^{-j 2 \pi f (t'-t)} dt'
\label{eq_th1} \\
& = \int_{t'} g_{\text{rx}}^*(t'-t) \bigg[\int_{\tau} \int_{\nu} h(\tau,\nu) \left\{ \sum_{n'=0}^{N-1}\sum_{m' = 0}^{M-1}   X[n',m'] g_{\text{tx}}(t'-\tau-n'T) e^{j2\pi m' \Delta f(t'-\tau-n'T)} \right\} \nonumber \\ & \hspace{10cm} e^{j2\pi\nu(t'-\tau)} d\tau d\nu \bigg] e^{-j 2 \pi f (t'-t)} dt'
\label{eq_th2} \\
& = \sum_{n'=0}^{N-1}\sum_{m' = 0}^{M-1}   X[n',m'] \bigg[\int_{\tau} \int_{\nu} h(\tau,\nu)  \bigg\{\int_{t'} g_{\text{rx}}^*(t'-t) g_{\text{tx}}(t'-\tau-n'T) e^{j2\pi m' \Delta f(t'-\tau-n'T)}\nonumber \\ & \hspace{9.5cm}e^{j2\pi\nu(t'-\tau)} e^{-j 2 \pi f (t'-t)} dt'  \bigg\} d\tau d\nu \bigg]
\label{eq_th3}
\end{align}
\hrule

\begin{align}
& H_{n,m}[n',m']  = \int_{\tau} \int_{\nu} h(\tau,\nu)  \bigg[\int_{t'} g_{\text{rx}}^*(t'-nT) g_{\text{tx}}(t'-\tau-n'T) e^{j2\pi m' \Delta f(t'-\tau-n'T)}e^{j2\pi\nu(t'-\tau)} \nonumber \\ & \hspace{11cm}e^{-j 2 \pi m \Delta f (t'-nT)} dt'  \bigg] d\tau d\nu
\label{eq_th4}\\
& = \int_{\tau} \int_{\nu} h(\tau,\nu)  \bigg[\int_{t''}g_{\text{rx}}^*(t''-(n-n')T+\tau) g_{\text{tx}}(t'') e^{j2\pi m' \Delta f t''} e^{j2\pi\nu(t''+n'T)}\nonumber \\ & \hspace{9.5cm} e^{-j 2 \pi m \Delta f (t''+(n-n')T+\tau)}   dt'' \bigg] d\tau d\nu
\label{eq_th5}\\
& = \int_{\tau} \int_{\nu} h(\tau,\nu)  \left[\int_{t''}g_{\text{rx}}^*(t''-(n-n')T+\tau) g_{\text{tx}}(t'') e^{-j 2\pi \left((m-m')\Delta f -\nu \right) \left(t''-(n-n')T+\tau \right)} dt'' \right]\nonumber \\ & \hspace{9cm} e^{j 2\pi (\nu+m'\Delta f) ((n-n')T-\tau)} e^{j 2\pi \nu n' T}  d\tau d\nu
\label{eq_th6}\\
& = \int_{\tau}\int_{\nu} \!\!h(\tau,\nu) A_{g_{rx},g_{tx}}\!((n-\!n')T-\!\!\tau,(m-\!m')\Delta f-\!\!\nu) e^{j 2\pi (\nu+m'\Delta f) ((n-n')T-\tau)} e^{j 2\pi \nu n' T}d\tau d\nu
\label{eq_th7}
\end{align}
\hrule

\begin{align}
y[k,l] & = \frac{1}{NM} \sum_{n=0}^{N-1} \sum_{m=0}^{M-1} H_{n,m}[n,m] \left[\sum_{k'=0}^{N-1}\sum_{l'=0}^{M-1} x[k',l'] e^{j2\pi\bigl(\frac{nk'}{N}-\frac{ml'}{M}\bigr)} \right] e^{-j2\pi\bigl(\frac{nk}{N}-\frac{ml}{M}\bigr)} \label{eq_rx1} \\
& = \frac{1}{NM} \sum_{k'=0}^{N-1}\sum_{l'=0}^{M-1} x[k',l']  \left[\sum_{n=0}^{N-1} \sum_{m=0}^{M-1} H_{n,m}[n,m] e^{-j2\pi nT \bigl(\frac{k-k'}{NT}\bigr)} e^{j2\pi m \Delta f \bigl(\frac{l-l'}{M\Delta f}\bigr)} \right] \label{eq_rx2}\\
& = \frac{1}{NM} \sum_{k'=0}^{N-1}\sum_{l'=0}^{M-1} x[k',l'] h_w[k-k', l-l'] \label{eq_rx3}.
\end{align}
\hrulefill
\begin{align}
h_w(\nu,\tau) & = \sum_{n=0}^{N-1} \sum_{m=0}^{M-1} \left[ \int_{\tau'} \int_{\nu'} h(\tau',\nu') {e^{j 2\pi \nu' nT}} e^{-j 2\pi(\nu' + m\Delta f)\tau'} d\tau' d\nu' \right] e^{-j2\pi nT \nu} e^{j2\pi m \Delta f \tau} 
\label{eq_htime1} \\
& = \int_{\tau'} \int_{\nu'} h(\tau',\nu') \left[ \sum_{n=0}^{N-1} \sum_{m=0}^{M-1} e^{-j 2\pi (\nu-\nu')nT } e^{j 2\pi (\tau-\tau')m\Delta f } \right] e^{-j 2\pi \tau' \nu'}  d\tau' d\nu'
\label{eq_htime2} \\
& = \int_{\tau'} \int_{\nu'} h(\tau',\nu') w(\nu-\nu',\tau-\tau') e^{-j 2\pi \tau' \nu'}  d\tau' d\nu'.
\label{eq_htime3}
\end{align}
%\hrulefill
\begin{align}
y[k,l]&= \frac{1}{\sqrt{NM}} \sum_{n=0}^{N-1} \sum_{m=0}^{M-1} \left[ \sum_{m'=0}^{M-1} H_{n,m}[n,m'] X[n,m'] + \sum_{m'=0}^{M-1} H_{n,m}[n-1,m'] X[n-1,m'] \right] \nonumber \\ & \hspace{11.5cm}e^{-j2\pi\bigl(\frac{nk}{N}-\frac{ml}{M}\bigr)}
\label{eq_rect1}\\
y_{\text{ici}}[k,l] & = \frac{1}{NM}  \sum_{n=0}^{N-1} \sum_{m=0}^{M-1} \sum_{m'=0}^{M-1} H_{n,m}[n,m'] \left[\sum_{k'=0}^{N-1}\sum_{l'=0}^{M-1} x[k',l'] e^{j2\pi\bigl(\frac{nk'}{N}-\frac{m'l'}{M}\bigr)}\right] e^{-j2\pi\bigl(\frac{nk}{N}-\frac{ml}{M}\bigr)}
\nonumber\\
& = \frac{1}{NM} \sum_{k'=0}^{N-1}\sum_{l'=0}^{M-1} x[k',l'] \left[\sum_{n=0}^{N-1} \sum_{m=0}^{M-1} \sum_{m'=0}^{M-1} H_{n,m}[n,m'] e^{-j2\pi n\bigl(\frac{k-k'}{N}\bigr)} e^{j2\pi\bigl(\frac{ml-m'l'}{M}\bigr)}  \right] \nonumber \\
& = \frac{1}{NM} \sum_{k'=0}^{N-1}\sum_{l'=0}^{M-1} x[k',l'] h^{\text{ici}}_{k,l}[k',l']
\label{eq_rect2}
\end{align}
%\vspace*{4pt}
%\end{figure*}
\hrule
%\begin{figure*}
\normalsize
\begin{align}
& h^{\text{ici}}_{k,l}[k',l']  = \sum_{n=0}^{N-1} \sum_{m=0}^{M-1} \sum_{m'=0}^{M-1} \bigg[\int_{\tau}\int_{\nu} h(\tau,\nu) A_{g_{rx},g_{tx}}(-\tau,(m-m')\Delta f-\nu)  e^{-j 2\pi (\nu+m'\Delta f) \tau} e^{j 2\pi \nu n T} \nonumber \\ & \hspace{9.7cm}d\tau d\nu\bigg]  e^{-j2\pi n\bigl(\frac{k-k'}{N}\bigr)} e^{j2\pi\bigl(\frac{ml-m'l'}{M}\bigr)} \label{eq_rect3} \\
&  = \frac{1}{M} \sum_{n=0}^{N-1} \sum_{m=0}^{M-1} \sum_{m'=0}^{M-1} \left[\sum_{i=1}^{P} h_i\! \! \! \! \sum_{p=0}^{M-1-l_{\tau_i}}  e^{-j2\pi((m-m')\Delta f-\nu_i)(p(T/M)+\tau_i)}  e^{-j 2\pi (\nu_i+m'\Delta f) \tau_i} e^{j 2\pi \nu_i n T}  \right] \nonumber \\ & \hspace{11cm} e^{-j2\pi n\bigl(\frac{k-k'}{N}\bigr)} e^{j2\pi\bigl(\frac{ml-m'l'}{M}\bigr)} \label{eq_rect4}\\
& =  \sum_{i=1}^{P} h_i \left[\sum_{n=0}^{N-1} e^{-j2\pi n\bigl(\frac{k-k'-k_{\nu_i} - \kappa_{\nu_i}}{N}\bigr)}\right] \bigg[\frac{1}{M} \sum_{p=0}^{M-1-l_{\tau_i}} e^{j 2 \pi \frac{p}{M} \left( \frac{k_{\nu_i}+\kappa_{\nu_i}}{N} \right) } \sum_{m=0}^{M-1} e^{-j 2 \pi (p+l_{\tau_i}-l) \frac{m}{M}} \nonumber \\ & \hspace{12cm} \sum_{m'=0}^{M-1} e^{j 2 \pi (p-l') \frac{m'}{M}}\bigg] \nonumber \\
& = \sum_{i=1}^{P}h_i \mathcal{G}^{ici}(\nu_i) \mathcal{F}^{ici}(\tau_i,\nu_i)
\label{eq_rect5}
\end{align}
\hrule
\begin{align}
\mathcal{F}^{ici}(\tau_i,\nu_i) &  =  {M}  \sum_{p=0}^{M-1-l_{\tau_i}} e^{j 2 \pi \frac{p}{M} \left( \frac{k_{\nu_i}+\kappa_{\nu_i}}{N} \right) } \delta([p+l_{\tau_i}-l]_M) \delta([p-l']_M)
\label{eq_rect6}
\end{align}
\begin{align}
& y_{\text{ici}}[k,l] \! =\!\! \frac{1}{N} \!\sum_{i=1}^{P} h_i\!
\!\left[\sum_{l'=0}^{M-1} \sum_{p=0}^{M-1-l_{\tau_i}}\! \!\!e^{j 2 \pi \frac{p}{M} \left( \frac{k_{\nu_i}+\kappa_{\nu_i}}{N} \right) } \!\delta([p+l_{\tau_i}-l]_M) \delta([p-l']_M) \! \sum_{k'=0}^{N-1} \mathcal{G}^{ici}(\nu_i)x[k',l']\right] \nonumber \\
%\label{eq_rect7} \\
& {\color{black} \approx} \frac{1}{N} \sum_{i=1}^{P}h_i \Bigg[ \sum_{p=0}^{M-1-l_{\tau_i}} e^{j 2 \pi \frac{p}{M} \left( \frac{k_{\nu_i}+\kappa_{\nu_i}}{N} \right) } \delta([p+l_{\tau_i}-l]_M) \sum_{q=-N_i}^{N_i} \left(\frac{e^{j {2\pi} (-q - \kappa_{\nu_i}) }-1}{e^{j \frac{2\pi}{N} (-q - \kappa_{\nu_i})}-1}\right) \nonumber \\ & \hspace{11cm}  x[[k - k_{\nu_i} + q]_N,p]\Bigg] 
\label{eq_rect8}
\end{align}
\begin{align}
 y_{\text{ici}}[k,l]  {\color{black} \approx}
\begin{cases}
\displaystyle\sum_{i=1}^{P} \displaystyle\sum_{q=-N_i}^{N_i} h_i  \left[\frac{1}{N}\beta_i(q)\right] e^{j2\pi \left(\frac{l-l_{\tau_i}}{M}\right) \left( \frac{k_{\nu_i}+\kappa_{\nu_i}}{N} \right)} x\left[[k - k_{\nu_i} +q]_N, [l - l_{\tau_i}]_M\right] &  l\geq l_{\tau_i}, \\
0 & {\text {otherwise.}}
\end{cases}
\label{eq_rect9}
\end{align}
% \begin{align}
% y_{\text{ici}}[k,l] & = 
% \begin{cases}
% \displaystyle\sum_{i=1}^{P} \displaystyle\sum_{k'=[k-k_{\nu_i}-N_i]_N}^{[k-k_{\nu_i}+N_i]_N} h_i  \left[\frac{1}{N}\beta_i(k,k')\right] e^{j\theta_i(l)} x\left[[k - k_{\nu_i} - k']_N, [l - l_{\tau_i}]_M\right] & {\text {if }} l\geq l_{\tau_i}, \\
% 0 & {\text {otherwise.}}
% \end{cases}
% \label{eq_rect9}
% \end{align}
%\hrule
\begin{align}
& y_{\text{isi}}[k,l]  = \frac{1}{NM}  \sum_{n=0}^{N-1} \sum_{m=0}^{M-1} \sum_{m'=0}^{M-1} H_{n,m}[n-1,m'] \left[\sum_{k'=0}^{N-1}\sum_{l'=0}^{M-1} x[k',l'] e^{j2\pi\bigl(\frac{(n-1)k'}{N}-\frac{m'l'}{M}\bigr)}\right] e^{-j2\pi\bigl(\frac{nk}{N}-\frac{ml}{M}\bigr)}
\nonumber\\
& = \frac{1}{NM} \sum_{k'=0}^{N-1}\sum_{l'=0}^{M-1} e^{-j2\pi \frac{k'}{N}} x[k',l'] \left[\sum_{n=0}^{N-1} \sum_{m=0}^{M-1} \sum_{m'=0}^{M-1} H_{n,m}[n-1,m'] e^{-j2\pi n\bigl(\frac{k-k'}{N}\bigr)} e^{j2\pi\bigl(\frac{ml-m'l'}{M}\bigr)}  \right] \nonumber \\
& = \frac{1}{NM} \sum_{k'=0}^{N-1}\sum_{l'=0}^{M-1} e^{-j2\pi \frac{k'}{N}} x[k',l'] h^{\text{isi}}_{k,l}[k',l']
\label{eq_rect10}
\end{align}
\begin{align}
h^{\text{isi}}_{k,l}[k',l'] & =  \sum_{i=1}^{P} h_i \left[\sum_{n=1}^{N-1} e^{-j2\pi n\bigl(\frac{k-k'-k_{\nu_i} - \kappa_{\nu_i}}{N}\bigr)}\right] \bigg[\frac{1}{M} \sum_{p=M-l_{\tau_i}}^{M-1} e^{j 2 \pi \left(\frac{p-M}{M}\right) \left( \frac{k_{\nu_i}+\kappa_{\nu_i}}{N} \right) } \nonumber \\ & \hspace{7cm} \sum_{m=0}^{M-1} e^{-j 2 \pi (p+l_{\tau_i}-l+M) \frac{m}{M}} \sum_{m'=0}^{M-1} e^{j 2 \pi (p-l') \frac{m'}{M}}\bigg] \nonumber \\
& = \sum_{i=1}^{P}h_i \mathcal{G}^{isi}(\nu_i) \mathcal{F}^{isi}(\tau_i,\nu_i)
\label{eq_rect11}
\end{align}
\hrule
{\color{black}
\begin{align}
& y_{\text{isi}}[k,l] = \frac{1}{N} \sum_{i=1}^{P} h_i \Bigg[\sum_{l'=0}^{M-1} \sum_{p=M-l_{\tau_i}}^{M-1} \!\!\!e^{j 2 \pi \left(\frac{p-M}{M}\right) \left( \frac{k_{\nu_i}+\kappa_{\nu_i}}{N} \right) } \delta([p+l_{\tau_i}-l]_M) \delta([p-l']_M) \cdot \nonumber \\ & \hspace{10.5cm} \sum_{k'=0}^{N-1} \mathcal{G}^{isi}(\nu_i) e^{-j2\pi \frac{k'}{N}} x[k',l']\Bigg] 
\label{eq_rect12} \\
&  = \frac{1}{N} \sum_{i=1}^{P}h_i \left[ \sum_{p=M-l_{\tau_i}}^{M-1} e^{j 2 \pi \left(\frac{p-M}{M}\right) \left( \frac{k_{\nu_i}+\kappa_{\nu_i}}{N} \right) } \delta([p+l_{\tau_i}-l]_M) \sum_{k'=0}^{N-1} \mathcal{G}^{isi}(\nu_i) e^{-j2\pi \frac{k'}{N}}x[k',p]\right] \nonumber \\ 
% & = \frac{1}{N} \sum_{i=1}^{P}h_i \left[ \sum_{p=M-l_{\tau_i}}^{M-1} e^{j 2 \pi \left(\frac{p-M}{M}\right) \left( \frac{k_{\nu_i}+\kappa_{\nu_i}}{N} \right) } \delta([p+l_{\tau_i}-l]_M) \right] \left[\sum_{q=-N_i}^{N_i} \left(\beta_i(q)-1\right)\right]  e^{-j2\pi \frac{[k - k_{\nu_i} + q]_N}{N}} \nonumber \\ & \hspace{12cm} x[[k - k_{\nu_i} + q]_N,p] 
% \label{eq_rect13}
& \approx \frac{1}{N} \sum_{i=1}^{P}h_i \Bigg[\sum_{p=M-l_{\tau_i}}^{M-1} e^{j 2 \pi \left(\frac{p-M}{M}\right) \left( \frac{k_{\nu_i}+\kappa_{\nu_i}}{N} \right) } \delta([p+l_{\tau_i}-l]_M)  \nonumber \\ & \hspace{10mm}  \left\{\sum_{q=-N_i}^{N_i}  \left(\beta_i(q)-1\right)  e^{-j2\pi \frac{[k - k_{\nu_i} + q]_N}{N}}  x[[k - k_{\nu_i} + q]_N,p] - \sum_{\mathclap{\substack{k'=0,\\ k'\neq [k - k_{\nu_i} + q]_N, q \in [-N_i, N_i]}}}^{N-1} e^{-j2\pi \frac{k'}{N}}x[k',p]\right\}  \Bigg]
\label{eq_new1} \\
&  \approx \frac{1}{N} \sum_{i=1}^{P}h_i \Bigg[ \sum_{p=M-l_{\tau_i}}^{M-1} e^{j 2 \pi \left(\frac{p-M}{M}\right) \left( \frac{k_{\nu_i}+\kappa_{\nu_i}}{N} \right) } \delta([p+l_{\tau_i}-l]_M) \sum_{q=-N_i}^{N_i} \left(\beta_i(q)-1\right)  e^{-j2\pi \frac{[k - k_{\nu_i} + q]_N}{N}} \nonumber \\ & \hspace{11cm} x[[k - k_{\nu_i} + q]_N,p] \Bigg]  \label{eq_new2}
\end{align}}
%\end{figure*}
%\begin{figure*}
\begin{align}
y_{\text{isi}}[k,l] &{\color{black}\approx} 
\begin{cases}
\displaystyle\sum_{i=1}^{P} \displaystyle\sum_{q=-N_i}^{N_i} h_i  \left[\frac{1}{N}\left(\beta_i(q)-1\right)\right] e^{-j2\pi \frac{[k - k_{\nu_i} + q]_N}{N}} e^{j2\pi \left(\frac{l-l_{\tau_i}}{M}\right) \left( \frac{k_{\nu_i}+\kappa_{\nu_i}}{N} \right)} \\ \hspace{6cm} x\left[[k - k_{\nu_i} +q]_N, [l - l_{\tau_i}]_M\right] &  l < l_{\tau_i}, \\
0 & {\text {otherwise.}}
\end{cases}
\label{eq_rect14}
\end{align}
%\end{figure*}
%\todo{Please check the highlighted parts for possible typos. $+M$ mod M seems to be unnecessary?}
\section{Proof of Proposition \ref{pro2}: OTFS Input--Output Relation in Delay--Doppler Domain for Ideal Pulses}
\label{app_pro2}
The received signal $y[k,l]$ for the ideal pulses, from (\ref{rx_sfft}) and (\ref{eq:receivedsig_Y}), can be written as
\begin{align}
y[k,l] & = \!\frac{1}{\sqrt{NM}} \!\sum_{n=0}^{N-1} \sum_{m=0}^{M-1}\! H_{n,m}[n,m] X[n,m] e^{-j2\pi\bigl(\frac{nk}{N}-\frac{ml}{M}\bigr)}.
\nonumber
\end{align}
By substituting the ISFFT equation from (\ref{iSFFT}), $y[k,l]$ can be expanded as in from (\ref{eq_rx1}) to (\ref{eq_rx3}). Here, $h_w[k-k', l-l']$ can be seen as the value of $h_w(\nu,\tau)$ sampled at $\nu = \frac{k-k'}{NT}, \tau = \frac{l-l'}{M\Delta f}$. The value of $h_w(\nu,\tau)$ can be obtained as from (\ref{eq_htime1}) to (\ref{eq_htime3}), by substituting $H_{n,m}[n,m]$ from (\ref{eq_h_freq}), which completes the proof.

\section{Proof of Theorem \ref{th_rect_dd}: OTFS Input--Output Relation in Delay--Doppler Domain for Rectangular Pulses}
\label{app_th_rect_dd}

We start with expanding $y[k,l]$ in (\ref{rx_sfft}) using the $Y[n,m]$ for rectangular pulses in (\ref{eq_freq_time_rect}) as in (\ref{eq_rect1}). We write $y[k,l]$ as
\begin{align*}
y[k,l] = y_{\text{ici}}[k,l] +  y_{\text{isi}}[k,l],
\end{align*} where $y_{\text{ici}}[k,l]$ and $y_{\text{isi}}[k,l]$ contains the first term and the second term of the summation in square brackets of (\ref{eq_rect1}), respectively. We analyze these ICI and ISI terms as below.

{\em Analysis of $y_{\text{ici}}[k,l]$:} The value of $y_{\text{ici}}[k,l]$ can be written as in (\ref{eq_rect2}) using the ISFFT of $X[n,m]$ given in (\ref{iSFFT}). Now, $h^{\text{ici}}_{k,l}[k',l']$ is expanded in (\ref{eq_rect3}) by using the $H_{n,m}[n,m']$ value in (\ref{eq_h_freq}). This can be further written as in (\ref{eq_rect4}) from the channel assumption in (\ref{del_dop}) and the cross-ambiguity function in (\ref{a_ici}). 

To write the expression in (\ref{eq_rect4}) to a simple form, let us separate the terms related to $n,m,m',$ and $p$. The terms related to $n$ are
\begin{align*}
\zeta_n & = e^{-j2\pi n\bigl(\frac{k-k'}{N}\bigr)} e^{j 2\pi \nu_i n T} \\
& = e^{-j2\pi n\bigl(\frac{k-k'-k_{\nu_i} - \kappa_{\nu_i}}{N}\bigr)}
\end{align*}
Here, we used the delay and Doppler taps defined in (\ref{delaytap}).
Similarly, the terms related to $m$ and $m'$ are 
\begin{align*}
\zeta_m & = e^{-j 2\pi m \Delta f (p(T/M)+\tau_i)} e^{j 2 \pi l \frac{m}{M}} \\
& = e^{-j 2 \pi (p+l_{\tau_i}-l) \frac{m}{M}}\\
\zeta_{m'} & = e^{j 2\pi m' \Delta f (p(T/M)+\tau_i)} e^{-j 2\pi m' \Delta f \tau_i} e^{-j 2 \pi l' \frac{m'}{M}}\\
& = e^{j 2 \pi (p-l') \frac{m'}{M}}.
\end{align*}
Finally, the terms related to $p$ are
\begin{align*}
\zeta_{p} & = e^{j 2\pi \nu_i (p(T/M)+\tau_i)} e^{-j 2\pi \nu_i \tau_i} \\
& = e^{j 2 \pi \frac{p}{M} \left( \frac{k_{\nu_i}+\kappa_{\nu_i}}{N} \right) }. 
\end{align*}
Therefore, from the above terms, the value of $h^{\text{ici}}_{k,l}[k',l']$ can be written as in (\ref{eq_rect5}), where $\mathcal{G}^{ici}(\nu_i)$ and $\mathcal{F}^{ici}(\tau_i,\nu_i)$ denote the terms in the first and second square brackets. 
The value of $\mathcal{G}^{ici}(\nu_i)$ is the same as the one studied in (\ref{g_val}) for ideal pulses case. 
Similar to the analysis of (\ref{f_eq}), $\mathcal{F}^{ici}(\tau_i,\nu_i)$ can be written as
in (\ref{eq_rect6}). Hence, by substituting (\ref{eq_rect5}) and (\ref{eq_rect6}) in (\ref{eq_rect3}), $y_{\text{ici}}[k,l]$ can be approximated as in (\ref{eq_rect8}). From (\ref{eq_rect8}), we can easily see that it is non-zero only if the following conditions satisfied:
\begin{align*}
p = [l-l_{\tau_i}]_M {\text{ and }} 0\leq p \leq M-1-l_{\tau_i}.
\end{align*}
These conditions are satisfied only if $l \geq l_{\tau_i}$ and $p = l-l_{\tau_i}$.
Finally, with the conditions on $l$ and $p$, $y_{\text{ici}}[k,l]$ can be obtained as in (\ref{eq_rect9}), where $\beta_i(q)$ is defined in (\ref{eq_beta1}).

{\em Analysis of $y_{\text{isi}}[k,l]$:} Similar to $y_{\text{ici}}[k,l]$ in (\ref{eq_rect2}), $y_{\text{isi}}[k,l]$ can be expanded as in (\ref{eq_rect10}). By substituting the value of $H_{n,m}[n-1,m']$ from (\ref{eq_h_freq}), cross-ambiguity function in (\ref{a_isi}), and similar analysis of separating terms for $h^{\text{ici}}_{k,l}[k',l']$, the value of $h^{\text{isi}}_{k,l}[k',l']$ can be obtained as in (\ref{eq_rect11}). Here, the summation $n$ starts from $1$ as the first symbol does not have previous symbol to experience ISI. Therefore, the value of $\mathcal{G}^{isi}(\nu_i)$ is equal to $\mathcal{G}^{ici}(\nu_i)-1$. {\color{black}Using the value of $\mathcal{G}^{isi}(\nu_i)$, $y_{\text{isi}}[k,l]$ can be approximated as in (\ref{eq_new1}). Further, the expression in (\ref{eq_new1}) can be approximated as in (\ref{eq_new2}) by neglecting the signals $x[k',p]$ for which $k'\neq [k - k_{\nu_i} + q]_N, q \in [-N_i, N_i]$, as their coefficients are very small ($1/N$) for practical values of $N$ (e.g., $N=64,128$).}  

Now, (\ref{eq_new2}) is non-zero only if the following conditions are satisfied:
\begin{align*}
p = [l-l_{\tau_i}]_M {\text{ and }} M-l_{\tau_i} \leq p \leq M-1.
\end{align*}
These conditions are satisfied only if $l < l_{\tau_i}$ and $p = l-l_{\tau_i}+M$. With these conditions, the value of $y_{\text{isi}}[k,l]$ is written in (\ref{eq_rect14}).

Finally, by combining (\ref{eq_rect9}) and (\ref{eq_rect14}), the value of $y[k,l]$ in (\ref{eq_rect1}) can be obtained as in (\ref{conv_eq_rect}), which completes the proof.

\end{document}